\documentclass[5p,4pt]{elsarticle}

\biboptions{sort,compress}
\usepackage{txfonts}
\usepackage{subfig,graphicx}
\usepackage{amssymb}
\usepackage{natbib}
\usepackage{threeparttable}
\usepackage[]{txfonts}
\usepackage{textcomp}
\usepackage[colorlinks=true,linkcolor=blue,citecolor=blue,
bookmarks=true,bookmarksopen=true,bookmarksnumbered=true,draft=false]{hyperref}
\usepackage[hyphenbreaks]{breakurl}

\newcommand{\eqb}{\begin{eqnarray}}
\newcommand{\eqe}{\end{eqnarray}}
\newcommand{\mel}{m_{\rm e}}

\newcommand{\tcr}{t_{\rm cr}}
\newcommand{\sth}{\sigma_{\rm T}}
\newcommand{\lpinj}{\ell_{\rm p}^{\rm inj}}
\newcommand{\leinj}{\ell_{\rm e}^{\rm inj}}
\newcommand{\gpmn}{\gamma_{\rm p,min}}
\newcommand{\gpmx}{\gamma_{\rm p,max}}
\newcommand{\gemn}{\gamma_{\rm e,min}}
\newcommand{\gemx}{\gamma_{\rm e,max}}
\newcommand{\tpesc}{t_{\rm p,esc}}
\newcommand{\teesc}{t_{\rm e,esc}}

\newcommand{\doppler}{\delta}

\newcommand{\fermi}{{\sl Fermi}-LAT}
\newcommand{\mrk}{Mrk~421}

\usepackage{graphicx}
\usepackage{epsfig} 
\usepackage{epstopdf}
\usepackage{color}

\journal{Astroparticle Physics}

\begin{document}

\begin{frontmatter}



\title{Time-dependent neutrino emission from \mrk \ during flares and predictions for IceCube}

\author[purdue]{Maria Petropoulou\corref{cor1}}
\ead{mpetropo@purdue.edu}

\author[tum]{Stefan Coenders\corref{cor2}}
\ead{stefan.coenders@tum.de}

\author[alberta]{Stavros Dimitrakoudis}
\ead{dimitrak@ualberta.ca}

\address[purdue]{Department of Physics and Astronomy, Purdue University, 525 Northwestern
Avenue, West Lafayette, IN 47907, USA}
\address[tum]{Technische Universit{\"a}t M{\"u}nchen, Boltzmannstr. 2 (Universecluster),  D-85748 Garching bei M{\"u}nchen, Germany}
\address[alberta]{Department of Physics, University of Alberta, Edmonton, Alberta T6G 2E1, Canada }
\cortext[cor1]{Principal corresponding author, Einstein Postdoctoral Fellow}
\cortext[cor2]{Corresponding author}

\begin{abstract}
Blazars, a subclass of active galactic nuclei, are prime candidate sources for
the high energy neutrinos recently detected by IceCube. Being one of the
brightest sources in the extragalactic X-ray and $\gamma$-ray sky as well as
one of the nearest blazars to Earth, Mrk~421 is an excellent source for
testing the scenario of the blazar-neutrino connection, especially during
flares where time-dependent neutrino searches may have a higher detection probability. Here, we
model the spectral energy distribution of Mrk~421 during a 13-day flare
in 2010 with unprecedented multi-wavelength coverage, and calculate the
respective neutrino flux. We find a correlation between the $>1$ PeV
neutrino and photon fluxes, in all energy bands.  
{Using typical IceCube through-going muon event
samples with good angular resolution and high statistics, we }
derive the mean event rate
above 100~TeV ($\sim0.57$~evt/yr) and show that it is 
comparable to that expected from a four-month quiescent period in 2009. Due to the short duration of the
flare, an accumulation of similar flares over several years would be necessary
to produce a meaningful signal for IceCube. To better assess this, we apply the
correlation between the neutrino and $\gamma$-ray fluxes to the 6.9~yr
\fermi \ light curve of Mrk~421. We find that the mean event count above 1 PeV
for the full IceCube detector livetime is $3.59\pm0.60$ ($2.73\pm0.38$)
$\nu_\mu+\bar{\nu}_\mu$  with (without) major flares  included in our
analysis. This estimate exceeds, within the uncertainties, the $95\%$
($90\%$) threshold value for the detection of one or more muon 
(anti-)neutrinos. Meanwhile, the most conservative scenario, where no correlation of
$\gamma$-rays and neutrinos is assumed, predicts $1.60\pm0.16$
$\nu_\mu+\bar{\nu}_\mu$ events. We conclude that a non-detection of
high-energy neutrinos by IceCube would probe the
neutrino/$\gamma$-ray flux correlation during major flares
or/and the hadronic contribution to the blazar emission.
\end{abstract}

\end{frontmatter}


\section{Introduction} 
Ground-based imaging Cherenkov observatories, such as H.E.S.S. \citep{hinton2004}, MAGIC \citep{lorenz2004}  and VERITAS \citep{holder2009},
in synergy with the {\sl Fermi}-Large Area Telescope (LAT) \citep{atwood2009},
have accumulated sufficient $\gamma$-ray data to convincingly prove that
blazars, a class of active galactic nuclei (AGN) whose jets point along our line of
sight, are efficient particle accelerators. It is commonly accepted that
particle acceleration, which, in principle, affects both electrons and protons,
takes place in an ``active'' region of the blazar jet, such as a standing
shockwave \citep{marscher_gear85, kazanas_ellison86} or in sites of
relativistic magnetic reconnection \citep{giannios2010,
giannios13, sironi_spitkovsky14}.  If this is the case, then it is expected that both leptonic and
hadronic emission processes will contribute to the production of the
multi-wavelength (MW) blazar emission (for a review, see \cite{boettcher10,
boettcher12}).  In a nutshell, in such scenarios the characteristic blazar
spectral energy distribution (SED) that shows two humps in a luminosity vs. frequency
 diagram  \citep{ulrichetal97,fossatietal98} is explained in terms of
electron synchrotron radiation (from radio up to UV/X-rays) and of
hadronic-related processes (from MeV to TeV $\gamma$-rays). The latter include
proton synchrotron radiation \citep{muecke_protheroe01, aharonian00,muecke03},
pion-related cascades \citep{mannheim93, mannheimbiermann92} and synchrotron
radiation of pion-produced pairs \citep{dpm14, cerrutietal14}.

Although theoretical models invoking high-energy protons have similar success
to leptonic models in fitting the SEDs
of blazars \citep{boettcherreimer13, mastetal13, cerrutietal14, weidinger2015,
diltz15, petroetal15}, there is still no direct evidence of proton acceleration
in blazar jets (for searches of correlation between AGN and ultra-high energy
cosmic-ray  events, see also \cite{tinyakov01, george08, pierre_auger08,
macolino12, pierre_auger15}). The ultimate proof for the existence of
high-energy protons in blazar jets can come only from the detection of
high-energy neutrinos \citep[e.g.][]{stecker91, halzenzas97}.

Neutrino production in AGN flares has been modeled in anticipation of
observations by previous neutrino telescopes, such as AMANDA
\citep{atoyandermer01}. It was also postulated by \cite{Mannheim1992fsrq} that
electron neutrinos produced during AGN flares could be observable by Fly's Eye, a cosmic-ray observatory. 
Both those models focused on flat spectrum radio quasars (FSRQ), and in particular 3C~279, as they were
assumed to have higher neutrino luminosities than BL Lacs. 
The inherent difficulty in
modeling ``orphan'' TeV flares (i.e. with no X-ray counterparts) with leptonic
synchrotron self-Compton (SSC) emission \citep{Bottcher2005} made such events an
enticing target for hadronic models; particularly the 2002 flare of 1ES
1959+650, which was investigated in that regard by Halzen and Hooper
\cite{halzen_hooper05} and Reimer et al. \cite{reimeretal05}. Soon afterwards,
Dermer et al. \cite{dermeretal07} presented a more detailed analytical
calculation of expected neutrino emission during FSRQ flares, taking photon-photon 
($\gamma\gamma$) absorption into account.

An accurate modeling of the neutrino emission in both quiescent and flaring
states of blazar emission, which acts complementary to model-independent
studies \citep[e.g.][]{halzenzas97,rachen98, doert12, fraijamarinelli15}, is
vital for the interpretation of observations by neutrino telescopes, especially
in the context of the recent discovery of astrophysical neutrino events in the
100~TeV-2~PeV energy range by IceCube \citep{icecube13,aartsen14}. In fact, 
the neutrino spectrum extends to the multi-PeV energy range thanks to 
the newest IceCube detection of a track-like neutrino event with energy 
significantly above 2~PeV \citep{multi_pev}.
The neutrino flux expected from a non-flaring blazar in the context of a
specific leptohadronic model for the blazar SED was recently presented in
\citep[][]{dpm14}--henceforth, DPM14, where the blazar \mrk \ was used  as a
testbed. This is one of the nearest  ($z=0.031$, \citep{deVaucouleurs1991}) and
brightest BL Lac sources in the very high energy (VHE; $E_{\gamma}>200$~GeV) sky
\citep[e.g.][]{Senturk2013} and extragalactic X-ray sky, which makes it an
ideal target of MW observing campaigns. In particular, the results of the 2009
MW campaign \citep{Abdo2011}, which covers approximately a four month
non-flaring period (``quiescence'') of \mrk \, were used in DPM14. The compiled
time-averaged SED was modeled using a numerical leptohadronic code
\citep{DMPR2012} that self-consistently treats the energy losses of all
radiating particles in the active region of the blazar. {Implications of our model regarding other individual blazars and the neutrino background emission from the whole BL Lac population were presented, respectively, in \citep{petroetal15} and \citep{padovanipetro15}.  }

In this paper, we expand upon the work of DPM14
by studying the neutrino emission from \mrk \ during a flaring 
period in both X- and $\gamma$-ray energy bands. To this end, we apply our
model to the 13-day flare of 2010 (MJD~55265-55277), having
an unprecedented MW (from radio up to TeV $\gamma$-rays) and
simultaneous (within 2-3 hours) coverage  \citep{aleksic15}.  This dataset, with its
wide coverage in energy and time domains, offers a unique opportunity to:
\begin{itemize}
    \item test the applicability of the model to an active state of blazar
    emission;
    \item study the evolution of the neutrino spectrum during a period of
    flaring activity and calculate the respective neutrino light curve;
    \item test possible correlations between the neutrino and photon fluxes in
    different energy bands (e.g. X-rays and $\gamma$-rays);
    \item calculate the neutrino flux from \mrk \ during a $\gamma$-ray
    flare and compare it against the one expected from a longer, but
    non-flaring period, i.e. in quiescence;
    \item make predictions about the cumulative number of neutrino events that
    IceCube should detect in $t$ years, after applying the photon-neutrino flux 
    correlations, if any, to the long-term $\gamma$-ray  (\fermi) \ light curve of \mrk.
\end{itemize}

By investigating the aforementioned issues, we plan to address 
the more general question of whether $\gamma$-ray flares determine the optimum
time window for high-energy neutrino detection from the nearby blazar \mrk.

This paper is structured as follows.  In \S\ref{model} we outline the adopted
theoretical framework and the numerical code. A description of the IceCube technical characteristics that enter the neutrino event rate 
calculation are presented in \S\ref{sec:eventrate}. The results of our model
application to the flaring period of \mrk \ in March 2010 are presented in
\S\ref{sec:flare2010}. We estimate the cumulative number of neutrino events from
\mrk \ in the five years of full IceCube livetime in \S\ref{sec:longterm},
proceed in \S\ref{discussion} with a discussion of our results and conclude in
\S\ref{summary}.

{For the calculation of the expected number of events by IceCube, we will focus on searches that use up-going muons {\cite{aartsen13diffuse, aartsen15, aartsen14ps}} rather than high-energy starting events (HESE) {\cite{aartsen13, aartsen14}}. Thanks to a better
reconstruction accuracy, larger statistics and lower energy thresholds, the up-going muon samples are better suited to searching for faint signals from potential neutrino point sources, as we will discuss in more detail in \S\ref{sec:eventrate}.}
For the required transformations between the reference systems of the blazar
and the observer, we have adopted a cosmology with $\Omega_{\rm m}=0.3$,
$\Omega_{\Lambda}=0.7$ and $H_0=70$ km s$^{-1}$ Mpc$^{-1}$. The redshift of
\mrk \ $z=0.031$ corresponds to a luminosity distance $d_{\rm L}=136$~Mpc.

\section{The model}
\label{model}

\subsection{Theoretical framework}
We adopt a one-zone leptohadronic model for the blazar emission, where the
low-energy emission of the blazar SED is attributed to synchrotron radiation of
relativistic electrons and the observed high-energy (GeV-TeV) emission is
assumed to have a photohadronic origin.

In particular, we assume that the region responsible for the blazar emission
can be described as a spherical blob of radius $R$, containing a tangled
magnetic field of strength $B$ and moving with a Doppler factor
$\doppler$. Protons and (primary) electrons are accelerated by some mechanism
whose details lie outside the immediate scope of this work. They are
subsequently injected  isotropically in the volume of the blob with a constant
rate, which is parametrized as  $\ell_{\rm i, inj}=L_{\rm i, inj} \sth / 4\pi R
m_{\rm i} c^3$, where $L_{\rm i, inj}$ denotes the injection luminosity as
measured in the rest frame of the emitting region, $\sth=6.65\times
10^{-25}$~cm$^2$  is the Thomson cross section and the subscript i denotes
protons or electrons (i=p,e).  These are assumed to escape from the emitting
region in a characteristic timescale, which is set equal to the photon crossing
time of the source, i.e. $\tpesc=\teesc=R/c$.  Their distributions at injection
are described as power-laws with index $s_{\rm i}$ in the energy range
$E_{\rm i, \min}=\gamma_{\rm i, \min}m_{\rm i}c^2$ to $E_{\rm i, \max}=\gamma_{\rm i, \max}m_{\rm i}c^2$.

Photons, neutrons and neutrinos complete the set of the five stable
populations, that are at work in the blazar emitting region. Pions ($\pi^{\pm}, \pi^0$), muons ($\mu^{\pm}$) and
kaons ($K^{\pm},K^0$) 
constitute the unstable particle populations, since they decay into
lighter particles.  The production of pions is a natural outcome of
photohadronic interactions between the relativistic protons and the internal
photons; the latter are predominantly synchrotron photons emitted by the
primary electrons.  The decay of charged pions results in the injection of
secondary relativistic electrons and positrons  ($\pi^{\pm}\rightarrow
\mu^{\pm}+\nu_{\mu}(\bar{\nu}_{\mu})$, $\mu^{\pm}\rightarrow
e^{\pm}+\bar{\nu}_{\mu}({\nu}_{\mu})+\nu_{\rm e}(\bar{\nu}_{\rm e})$), whose
synchrotron emission emerges in the GeV-TeV regime, for a certain range of
parameter values.  Neutral pions  decay into VHE $\gamma$-rays (e.g.
$E_{\gamma}\sim10$~PeV, for a parent proton with energy $E_{\rm p}=100$~PeV),
and those are, in turn, susceptible to $\gamma \gamma$
absorption and can initiate an electromagnetic (or hadronic) cascade
\citep{mannheim91, mannheim93}. As SSC emission from primary electrons may also emerge in the GeV-TeV
energy band, the observed $\gamma$-ray emission can be totally or partially
explained by photohadronic processes, depending on the specifics of individual
sources \citep{petroetal15}. 

Besides neutrinos produced by photohadronic interactions between protons and
photons in the emission region of \mrk \,, an additional component (cosmogenic
neutrinos) may emerge from the interaction of escaping protons from the
source with the background radiation fields, such as the extragalactic
background light (EBL) \citep{stecker68}. In this study, we will neglect the
cosmogenic neutrino component, for reasons to be discussed in \S\ref{discussion}.

\subsection{Numerical framework}
The interplay of the processes governing the evolution of the energy
distributions of the five stable particle populations is formulated with a set
of five time-dependent, energy-conserving kinetic equations.  To simultaneously
solve the coupled kinetic equations for all particle types we use the
time-dependent code described in \cite{DMPR2012}. Photopion interactions are
modeled using the results of the Monte Carlo event generator \textsc{sophia}
\citep{SOPHIA2000}, while Bethe--Heitler pair production is similarly modeled
with the Monte Carlo results of Protheroe and Johnson \cite{Protheroe1996} and
Mastichiadis et al. \cite{mastetal05}. Details of the numerical treatment of short-lived particles (i.e., $\mu^{\pm}$, $\pi^{\pm}$, $\pi^0$, $K^{\pm}$ and $K^0$), which are not modeled with kinetic equations, can be found in \citep{dpm14, petroetal14}.  We finally note that for the  range of parameter values used in this study, the effect of synchrotron cooling for these unstable
particles is negligible.

\section{Neutrino point source detection with IceCube}
\label{sec:eventrate}
\begin{figure}
    \centering
    \includegraphics[width=0.47\textwidth,]{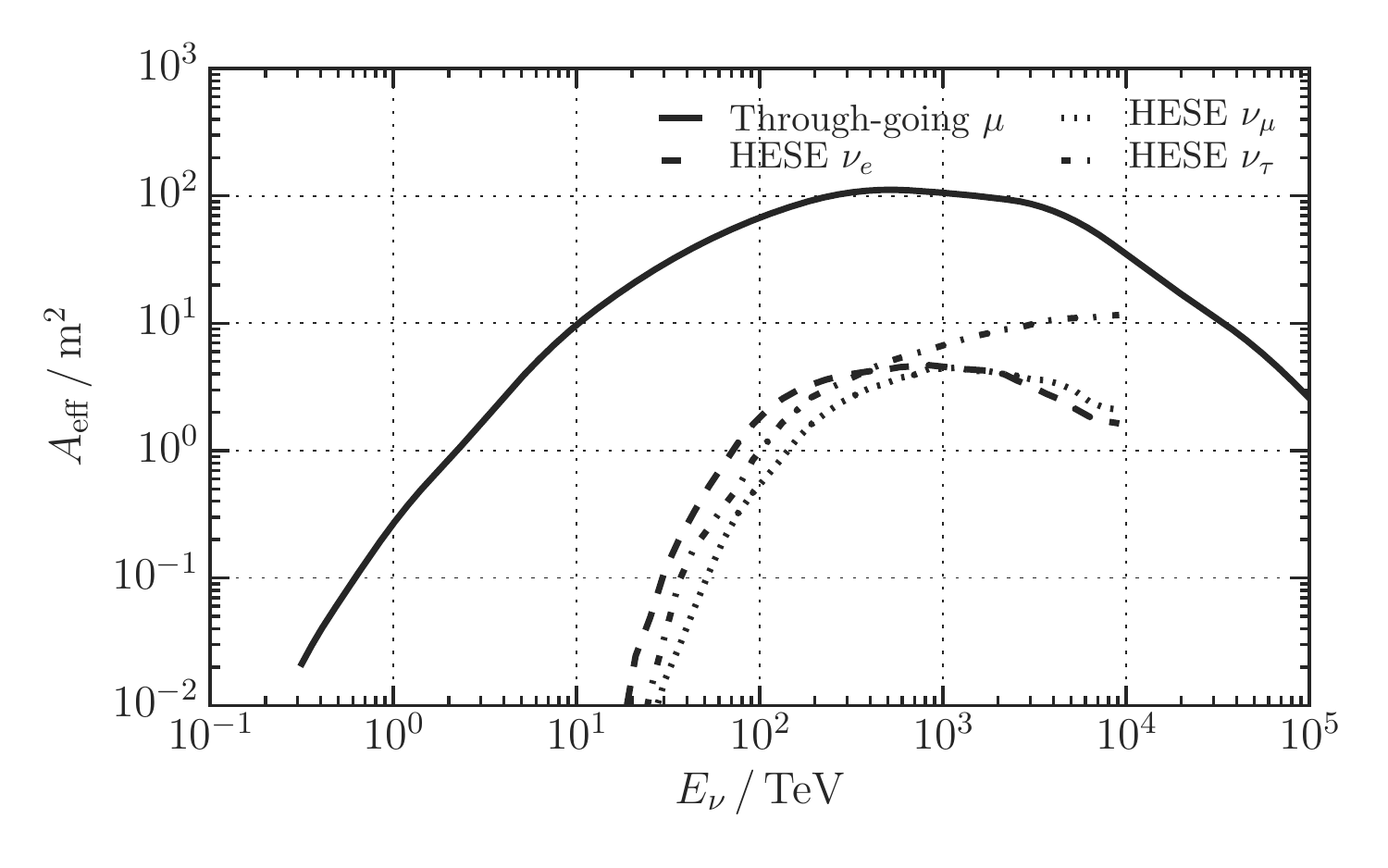}
    \caption{Effective area of IceCube $A_{\rm eff}$ at the position
             of \mrk \ (Ra: 166.07\textdegree, Dec: 38.19\textdegree) with
             respect to the primary neutrino energy $E_\nu$. The effective area
             is shown for typical up-going muon analysis (solid line) {and
             compared to that of the high-energy starting event (HESE) analysis for $\nu_{\rm e}$ (dashed line), $\nu_\mu$ (dotted line),
             and $\nu_\tau$ (dashed-dotted line).} Data are adopted
             from~\citep{aartsen13,aartsen14ps}.}
    \label{fig:IceCube_effA}
\end{figure}

The sensitivity of neutrino telescopes to a neutrino point source is limited
by vast backgrounds of $\mu^\pm$ produced in extensive air showers or by
charged-current (CC) interactions of atmospheric $\nu_\mu+\bar{\nu}_\mu$. 
Still, neutrino telescopes can cope with these
backgrounds with focused searches, using track-like events of $\mu^\pm$
penetrating the detector \cite{aartsen14ps, antares_ps}, for the following
reasons: (i) the angular reconstruction accuracy reduces the background to a
small part of the sky, while the expected mean background rate can be
effectively calculated using off-source regions; (ii) track-like events can
travel long distances before being detected, thus yielding a large collection
volume which increases with energy, yielding an effective area 10-100 times
larger than that for starting events \cite{icecube13} {(see also Fig.~\ref{fig:IceCube_effA});}
(iii) $\mu^\pm$ created in CC $\nu_\mu$
interactions are closely correlated to the parent $\nu$ direction above TeV
energies (there, the $90\%$ limit on the direction is well below 1\textdegree \,).
However, in track-like events, contrary to the cascade events, the information
of the $\mu$ and parent $\nu_\mu$ energies is partially lost, since only a small
fraction of the energy deposition along the track is observed.

{Even though the aforementioned searches are restricted to single-flavor neutrinos ($\nu_\mu+\bar{\nu}_\mu$), 
only muons created in CC interactions are reconstructed accurately enough to allow for a robust association of a neutrino with an astrophysical point source.
Furthermore, the position of \mrk \ in the northern sky (Dec:~$38.19^\circ$) coincides with that region in the sky where  IceCube is most
efficient in detecting muon flavored neutrinos (in the energy range of 100~TeV to a couple~PeV).  A small additional
component can arise from $\nu_\tau$ CC interactions followed by the sub-sequent decay of the $\tau$-lepton into $\mu\nu_\mu\nu_\tau$ with branching
ratio of $17\%$\footnote{Due to the three-body decay of the $\tau$ lepton, the energy
of the final $\mu$ will be lower than that of a $\mu$ produced by CC interactions of muon neutrinos.}.
In addition, regeneration of $\nu_\tau$  occurs during propagation within the Earth \cite{Bugaev2003},
increasing the flux at lower energies. The effect of $\tau$ neutrinos on our calculated rates is expected to be $\lesssim 10\%$  (compare \cite{abbasi2010}), i.e. smaller than other uncertainties considered in this work. Thus, in what follows we consider only  the muon component of the neutrino signal.}

\begin{figure}
     \centering
     \includegraphics[width=0.47\textwidth]{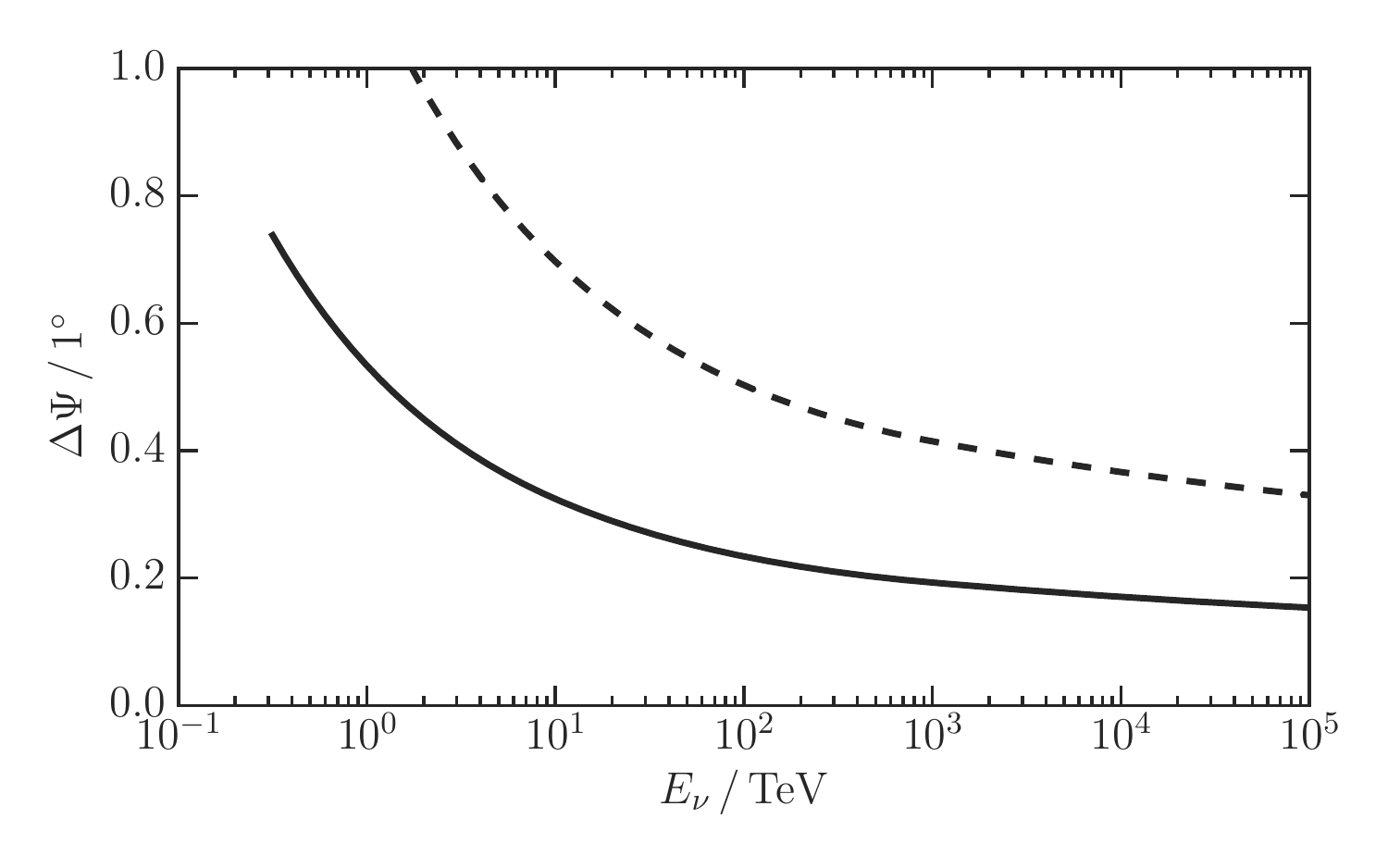}
    \caption{Energy dependence of the median angular resolution of IceCube $\Delta \Psi$ 
             (solid line) and of the estimated $90\%$ upper limit (dashed line) assuming a
             Gaussian distribution of the angular uncertainty. Data are
             adopted from \citep{aartsen14ps}.}
    \label{fig:IceCube_mrs}
\end{figure}

Taking into account the performance of the completed IceCube detector for
up-going track-like events \citep{aartsen14ps}, the expected (mean) number of
(anti-)neutrinos is then calculated by
\eqb
\label{eq0}
    N_{\nu} =  T \int_{E_{\nu,\min}}^{E_{\nu, \max}} \!\! dE_\nu
                \int_{\Delta\Omega(E_\nu)}\!\!\!\!\!\!\!\!\! {\rm d}\Omega \ A_{\rm eff}(E_\nu,
                \vec{x}\,)
                \sum_i
                \frac{\partial^2 F_{\nu,i}}{\partial\Omega\partial E_\nu},
\eqe
where $E_{\nu, \min}=100$~GeV, $E_{\nu,\max}=100$~PeV,
$\partial^2F_{\nu,i}\,/\,\partial\Omega\partial E_\nu$ is the incident muon
neutrino flux for different flux components $i$, and $\Delta\Omega$ is the
observation window around the source position $\vec{x}$. Equation (\ref{eq0})
shows that the mean number of events measured by IceCube depends mainly on:
\begin{enumerate} 
    \item the energy range of the flux; neutrino telescopes such as IceCube
    start to observe neutrinos in point source searches at TeV energies. At
    higher energies, the increasing cross-section
    $\sigma_\mathrm{CC}\left(\nu_\mu+X\rightarrow\mu+Y\right)$ enhances the
    effective area $A_{\rm eff}$ for neutrinos (see
    Fig.~\ref{fig:IceCube_effA}).  
    \item the point of observation; for the position of \mrk \ on the sky  (Ra:
    166.07\textdegree, Dec: 38.19\textdegree), the effective area
    $A_\mathrm{eff}$ of IceCube increases up until $1$~PeV before Earth
    absorption becomes dominant (see Fig.~\ref{fig:IceCube_effA}). 
    \item the contamination of the signal; for the northern sky, atmospheric
    $\bar{\nu}_\mu+\nu_\mu$ form an irreducible background over the high-energy
    signal. This will be exemplified later in \S\ref{subsec:neutrino} for the case of \mrk.
    \item the integration time of observation $T$.  
\end{enumerate} 
In what follows, we assume $90\%$ of all $\nu_\mu+\bar{\nu}_\mu$ events to be
reconstructed within 1\textdegree,  neglecting the energy dependence of 
the IceCube median angular resolution, $\Delta \Psi$. Given that resolution, which is
 shown in Fig.~\ref{fig:IceCube_mrs}, our choice provides a conservative
estimate of the event number.

The incident neutrino flux $\partial^2 F_{\nu,i}\,/\,\partial\Omega\partial E_\nu$
is primarily composed of three components.  The first component is the
signal from \mrk. \ Its flux is reduced by $10\%$ of events that are
reconstructed outside of the observation window.
The other two components are related to
the background neutrino emission  which, in turn, consists of (i) the
conventional atmospheric flux with a soft spectrum of approximately $\sim
E^{-3.7}$ \citep{Honda2007} and  (ii) the astrophysical flux, as measured by
IceCube with a spectral index $\sim E_\nu^{-2.3}$~\citep{aartsen14}. Here,
we treat this component as purely isotropic, thus, forming an additional
background at high energies for searches of point-like neutrino sources.
Within the window size $\Delta \Omega$, which is small compared to the variations of
reconstruction accuracy and effective area, the neutrino flux and effective
area are assumed to be constant. Thus, the integral over the solid angle in eq.~(\ref{eq0}) reduces to a 
constant $\Delta\Omega$. An additional prompt neutrino component is neglected, as it is
sub-dominant to the atmospheric or diffuse flux, both in the low- and
high-energy regimes, respectively.
\section{The 13-day flare of 2010}
\label{sec:flare2010}
Here, we present our results on the photon and neutrino emission from the
blazar \mrk \ during the 13-day flaring event of 2010, focusing on the expected neutrino event rate from that flare and on the
calculation of the IceCube sensitivity for \mrk. We then extrapolate our
findings using the long-term  ($\sim 6.9$~yr) \fermi \ $\gamma$-ray light curve
of \mrk \ and make predictions about the cumulative number of events that
IceCube should detect in the following years of its operation.  

The SEDs for the period MJD~$55265-55277$ were modeled by varying six out of the eleven free model
parameters (see Table~\ref{table0}), while the rest of them were kept fixed to
the following values: $B=5$~G, $R=3.2\times 10^{15}$~cm, $\gemn=100$, $\gpmn=1$
and $s_{\rm p}=1.2$. The values we chose for each of the six varying parameters may not necessarily correspond to the best possible fit for each day, as would be expressed by a $\chi^2$ minimum. Nevertheless, a good agreement between the model and the MW data is obtained for the whole duration of the flare. As the neutrino spectra are not sensitive to small changes in the model parameter values, the derived neutrino rates are robust.
This is the same approach as the one followed in DPM14 and \cite{petroetal15}. 

To minimize the computing time required for modeling the
13-day flaring activity we approximated the flaring period by a series of 13 steady-state snapshots. The
individual daily SEDs were numerically calculated for different values of the
six varying parameters. These were used as initial conditions for the numerical
calculation of the final steady state of the system (for continuous parameter
variations in time, see e.g. \citep{mastetal13}).  We note that our
approximation is valid as long as the typical time for reaching a steady-state
is less than the time interval  between two successive snapshots (1 day).
Indeed, a steady-state in our simulations  was typically achieved within $\sim 3 t_{\rm cr}= 3\times 10^5\, {\rm s}  \left(R/10^{15} {\rm cm}\right) < 2\times10^6 \, {\rm s} \left(\doppler/20\right)\left(\delta T/1 \, {\rm d}\right)$. 
We finally note that for the adopted parameter values the 
emission region is optically thin to photopion production (see also \S\ref{discussion}) with implications on the blazar
energetics, which have been discussed in \citep{petroetal15,padovanipetro15}.  
\subsection{Photon emission}
\begin{figure*}
 \centering
\includegraphics[width=0.8\textwidth]{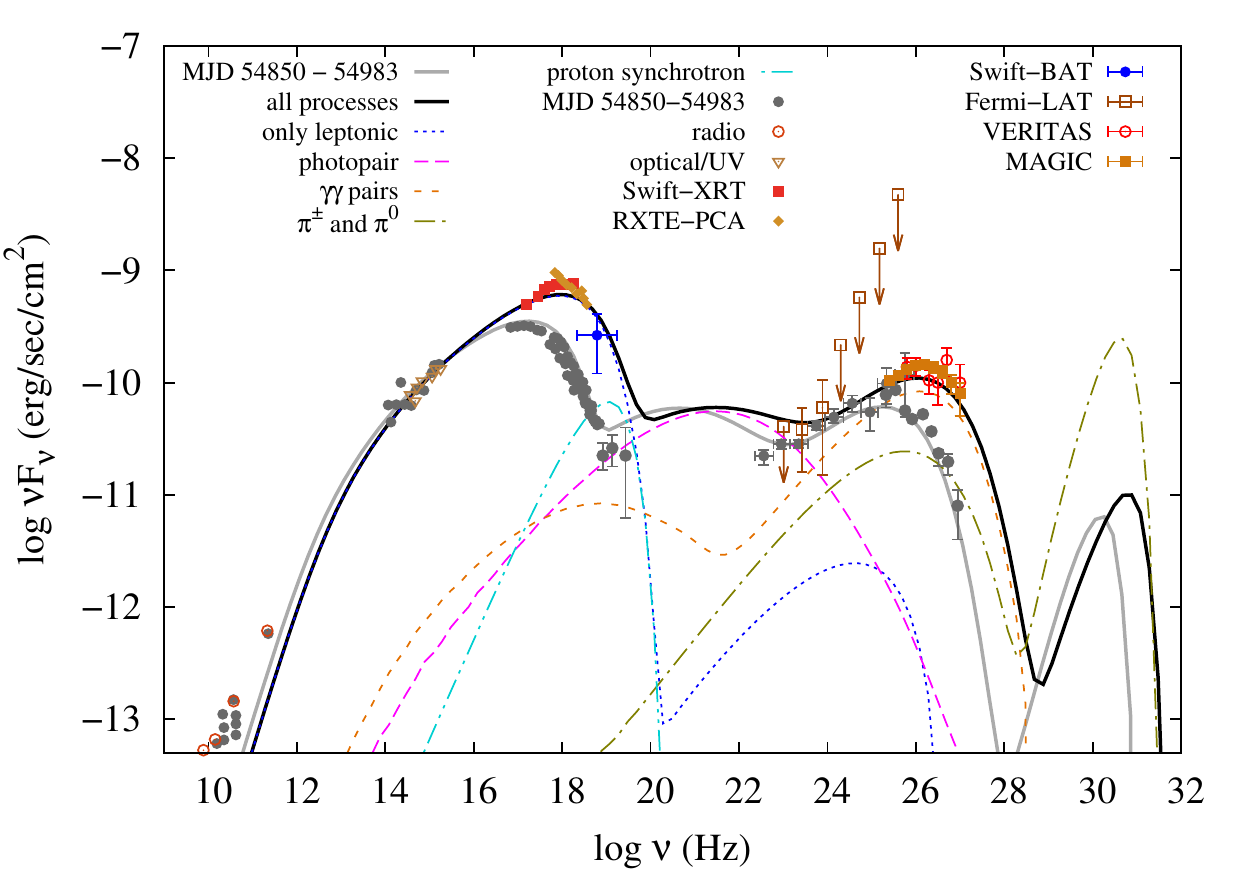}
 \includegraphics[width=0.8\textwidth]{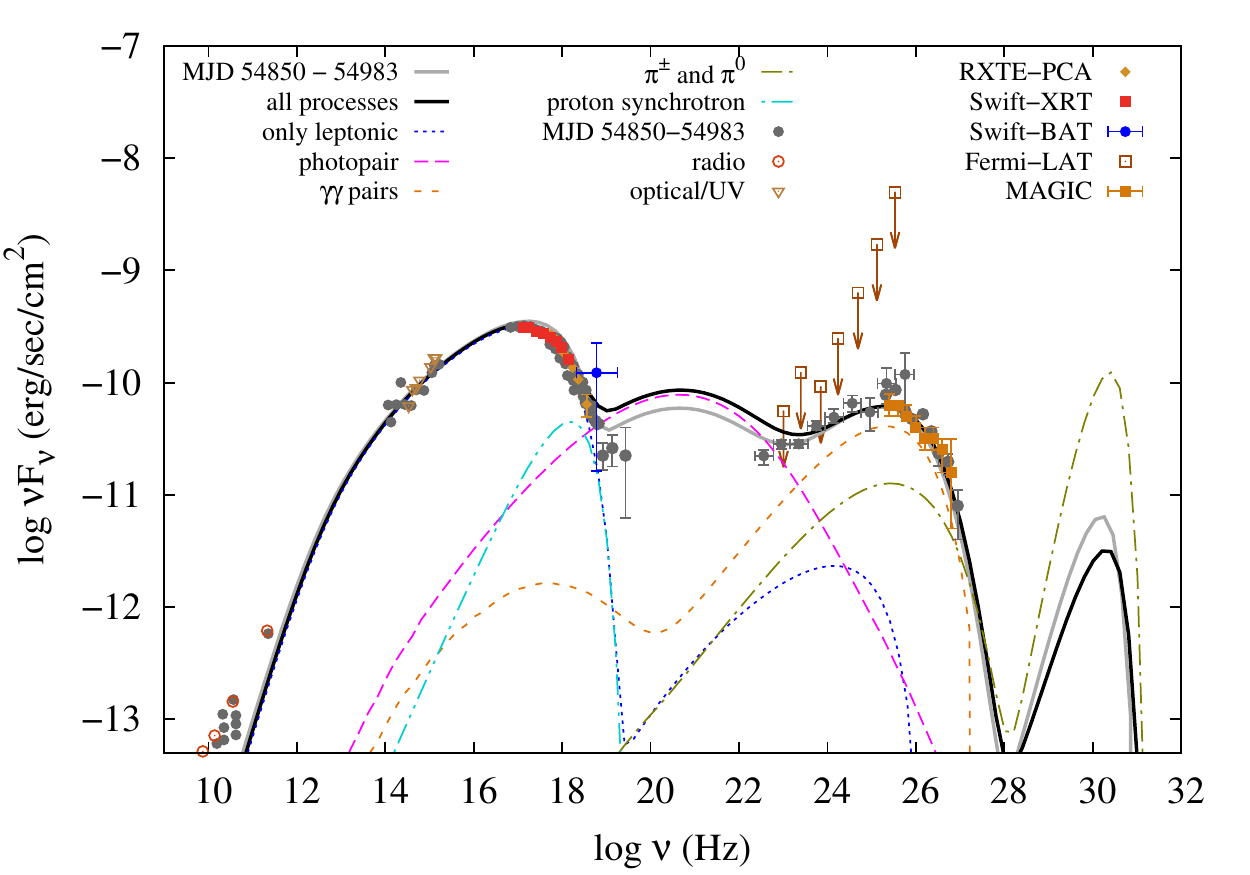}
 \caption{Simultaneous multi-wavelength SED of \mrk \ on  MJD~55265 (top
          panel) and  MJD~55277 (bottom panel).   
          Different symbols denote the various instruments used to collect  the data,
          and their meaning is given in the legends. All data-points are from
          \citep{aleksic15}. The grey circles depict the  time-averaged SED of
          \mrk \ over the period MJD~54850-54983 \citep{Abdo2011}. This  is a
          good representation of the blazar  non-flaring (quiescent)
          emission. The model-derived spectra that fit the daily SEDs are
          plotted with  black thick lines. The grey thick lines are a fit to
          the quiescent emission. Different types of lines are used to
          present the spectra from different emission processes: proton synchrotron radiation
          (light blue dashed double-dotted line),
          (primary) electron synchrotron and SSC emission (blue dotted line), synchrotron radiation from Bethe-Heitler pairs (magenta long-dashed line), 
          synchrotron radiation  of pairs from $\pi^{\pm}, \mu^{\pm}$ decays and $\gamma$-rays from $\pi^0$ decays (gold dashed-dotted line), synchrotron radiation of pairs from $\gamma \gamma$ absorption (orange short-dashed line).
           }
 \label{fig1}
\end{figure*}

The observed SEDs of \mrk \ for the first (MJD 55265) and last (MJD 55277)  days of the MW campaign are shown in Fig.~\ref{fig1}. To facilitate a comparison, the time-averaged SED over the period MJD~54850-54983 \citep{Abdo2011}, a good
representation of the blazar quiescent emission, has been included in the plot  (grey points). 
The model-derived photon spectra for the two days of the 2010 flare and of the 2009 quiescent period are plotted with thick  black and grey lines, respectively. The spectra produced by different emission processes are overplotted with different types
of lines (for details, see figure caption). The model SEDs for the rest of
the days are summarized in Fig.~\ref{fig2} of \ref{appenA}.
We note that the VHE ($>$200 GeV) observations have been already corrected for
absorption on the EBL in both \cite{aleksic15} and \cite{Abdo2011}. In other
words, the VHE $\gamma$-ray spectra shown in Figs.~\ref{fig1} and
\ref{fig2} are de-absorbed, and  the model-derived photon spectra
take  into account only the intrinsic $\gamma \gamma$ absorption. This also
explains the presence of the $\pi^0$ $\gamma$-ray bump at $\sim5-10$~PeV,
which otherwise  would be attenuated by the EBL. Figures~\ref{fig1} and \ref{fig2} show that the leptohadronic model
provides an overall good description of the data for the 13 consecutive days of
the flare. 

The \fermi \ observations at $\sim 400$~MeV are the
more constraining for our model, since for the adopted
parameter values the latter predicts  a luminous Bethe-Heitler component from hard X-rays to soft $\gamma$-rays (magenta long-dashed lines) \citep{petromast15}.
The Bethe-Heitler component, which is explained as synchrotron radiation of secondary electron-positron pairs 
produced via the Bethe-Heitler process, is a distinct 
feature to be constrained  with current, e.g. IBIS/INTEGRAL \citep{ubertini2003} and future, e.g. PANGU \citep{wu2014}, $\gamma$-ray satellites operating in the 1-100~MeV energy range. In addition, the Bethe-Heitler emission is expected to be {highly polarized and, as such, its modeling constitutes a prediction that may be tested by future $\gamma$-ray polarimeters,} such as ASTROGAM\footnote{\url{http://astrogam.iaps.inaf.it/scientific_instrument.html}} and AdEPT \citep{hunter2014}.
A comparison of the various emission components between the first and last days of the 13-day flare gives insight into the interplay of the different emission processes.  Figure~\ref{fig1} shows that, in both cases, the primary leptonic SSC component (blue dotted-line) peaking at $\sim 20$~GeV is sub-dominant compared to the emission from secondary pairs. In fact, the observed $\gamma$-ray emission in the range 2~GeV-2~TeV can be totally explained in the current model as synchrotron radiation from secondary pairs. These are the by-product of $\pi^{\pm}, \mu^{\pm}$ decays (gold dashed-dotted lines) and $\gamma \gamma$ absorption (orange short-dashed lines). It is noteworthy that the leptohadronic model shown here may degenerate into a leptonic one, with the SSC component dominating in $\gamma$-rays, simply by decreasing the injection luminosity in high-energy protons. As we discuss later in \S\ref{discussion},  IceCube will soon be in a position to constrain the contribution of hadronic-related processes to 
the $\gamma$-ray emission of \mrk.

The attenuation of VHE $\gamma$-rays produced via $\pi^0$ decays with energies $\sim 10$~PeV  leads to injection of high-energy pairs, whose synchrotron 
emission, for the adopted parameter values, peaks at $\sim 0.1-1$~TeV; this appears as a high-energy bump in the spectra shown with orange short-dashed lines (see Fig.~\ref{fig1}). A fraction of the (sub)TeV radiation is, in turn, attenuated leading to the production of the lower-energy bump of the spectrum that is plotted  with orange short-dashed lines in Fig.~\ref{fig1}. We note that the full width at half maximum (FWHM) of the two bumps is also related to the FWHM of the respective parent $\gamma$-ray spectra.  The (unattenuated) flux produced via photomeson processes (gold dashed-dotted lines) is highest at the start of the 13-day flare (left panel in Fig.~\ref{fig1}) and decreases towards the end of the flare (right panel in Fig.~\ref{fig1}). Since a fraction of the $\gamma$-ray flux is internally attenuated, the gradual $\gamma$-ray flux decrease over the 13-day period will be also reflected in the synchrotron emission of pairs produced by $\gamma \gamma$ absorption. Indeed, the peak flux of the lower 
energy bump in the synchrotron 
spectrum of $\gamma \gamma$ pairs (orange short-dashed lines) has decreased since the start of the 13-day flare (see both panels in Fig.~\ref{fig1}). 

Finally, the proton synchrotron spectrum (light blue dashed double-dotted lines) is the least variable  component, in terms of flux, during the 13-day flare. The peak energy of the spectrum has, however, decreased by approximately a factor of 7 between the start and end of the 13-day flare. 
The proton synchrotron spectrum peaks in hard X-rays, i.e. $5-35$~keV, and may
have a non-negligible contribution to the observed hard X-ray flux in other,
even  more extreme, flares  (see also MJD~55271 in Fig.~\ref{fig2}).  A great
example is the major MW flare of April 2013 \citep{balokovic13, cortina13,
hovatta13, paneque13}, where the fractional  variability in hard X-rays
(3-79~keV) as measured by {\sl NuSTAR} was found to be $0.790\pm0.001$
\citep{paliya15}, in contrast to $0.42\pm 0.12$ that was measured with BAT
(15-50~keV) during the 13-day flare \citep{aleksic15}. Regardless, the detailed
modeling of such an extreme flare across the MW spectrum as well as its
implications for the current model will be the subject of a subsequent paper.

The parameter values used in modeling the 13 daily SEDs are summarized in Table~\ref{table0}. 
Inspection of the table shows that no major variations of the model parameters were required for explaining the SEDs.  In all cases, the parameters
change by a factor of  $\sim 2.5$ at maximum with respect to their
time-averaged values. We find that an anti-correlation between $\gamma_{\rm p,
\max}$ and $\ell_{\rm p}$  is required to explain the data. This is an outcome
of the adopted flat proton spectrum $p<2$; a simultaneous increase of  both
$\gamma_{\rm p, \max}$ and $\ell_{\rm p}$, would lead to larger variations of
the $\gamma$-ray flux  than  what is actually observed. For the adopted
parameters, the electron distribution is modified by synchrotron cooling and
the peak synchrotron flux is therefore produced by electrons with Lorentz
factors equal to the cooling Lorentz factor. {This is defined as the
Lorentz factor where the synchrotron cooling time scale $6\pi
\mel c / \sth B^2 \gamma_{\rm e}$ equals the dynamical one $\tcr$}. Thus,
changes of $\gamma_{{\rm e}, \max}$ alone do not have a direct effect on the
peak synchrotron flux, which, in turn, explains the absence of correlation
between $\gamma_{\rm e, \max}$ and $\ell_{\rm e}$.
\begin{figure}
 \centering
 \includegraphics[width=0.48\textwidth]{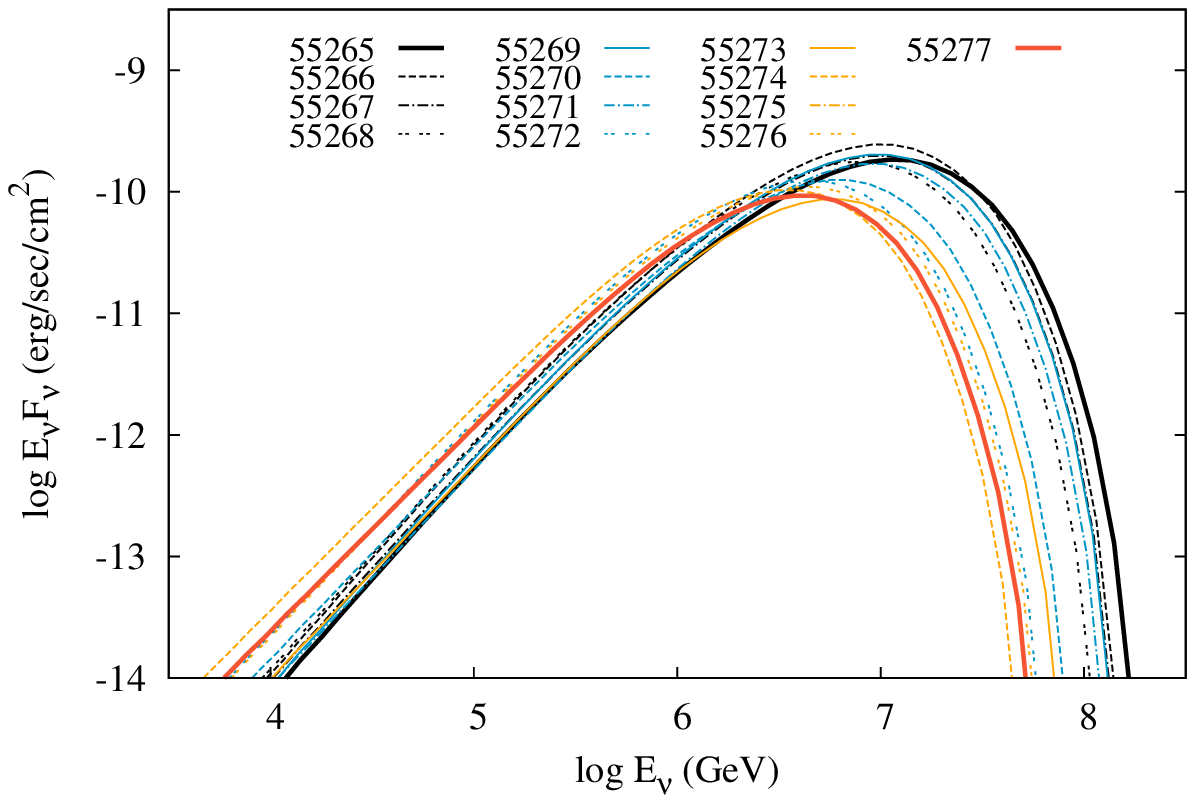} 
 \caption{All-flavor neutrino ($\nu + \bar{\nu}$) fluxes derived by the model
          for the period MJD~55265-55277. 
          } 
 \label{fig8}
\end{figure}
\begin{figure}
\centering
 \includegraphics[width=0.5\textwidth]{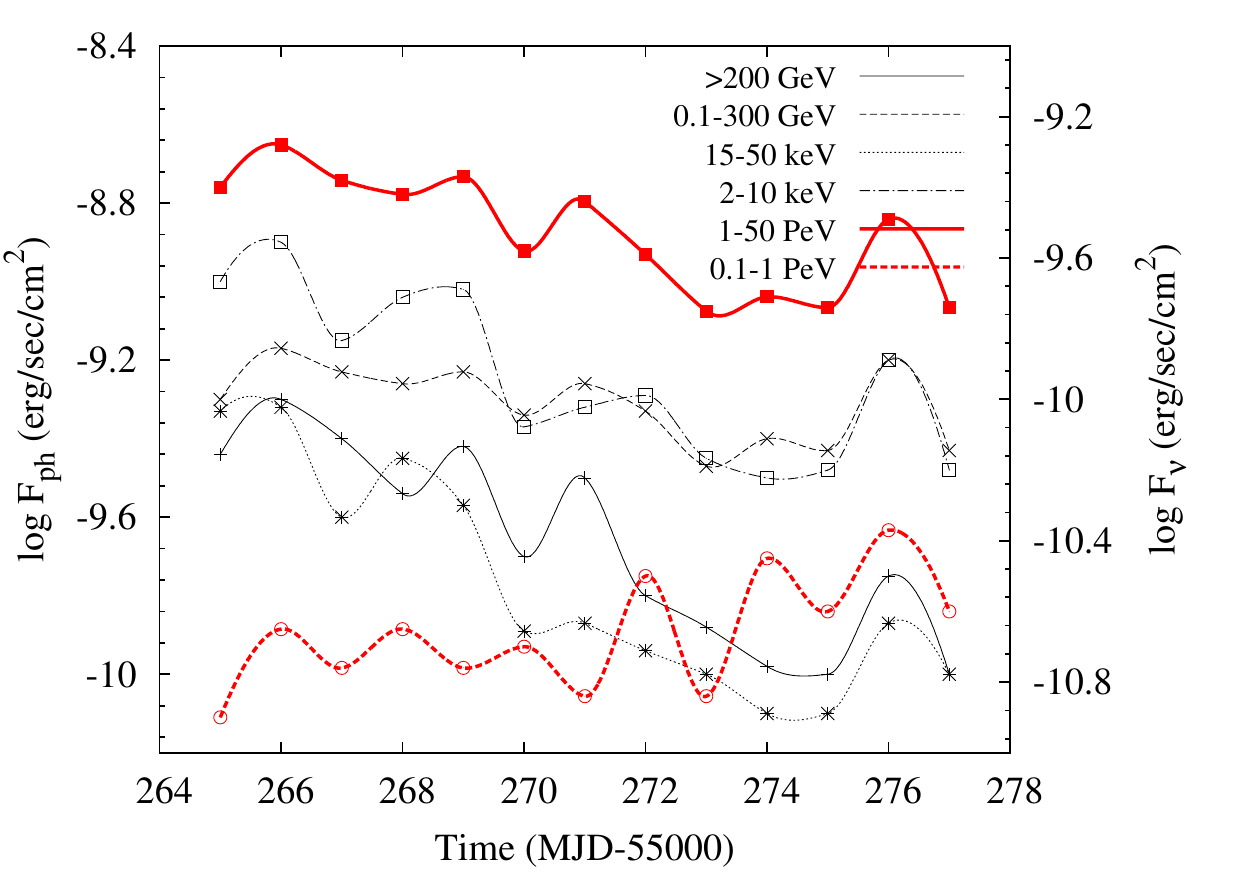}
 \caption{Model-derived light curves of \mrk \ covering the period
          MJD~55265-55277. Symbols denote the daily fluxes photons 
          ($F_{\rm ph}$) and neutrinos ($F_\nu$), while
          continuous lines are the result of interpolation.  Photon light
          curves (black lines) are calculated at four energy bands (see inset
          legend). The all-flavor neutrino ($\nu + \bar{\nu}$) light curves at
          1-50~PeV and 0.1-1~PeV  energy bands are also plotted with red solid
          and dashed lines, respectively. In all cases, the smooth curves are the result of
          interpolation.  }
 \label{fig9}
\end{figure}
\subsection{Neutrino emission}
\label{subsec:neutrino}
The daily all-flavor neutrino ($\nu + \bar{\nu}$) spectra are presented in
Fig.~\ref{fig8}, where thick lines are used for displaying the neutrino emission at
the beginning and the end of the 13-day flare. 
The low-energy ($<$1~PeV) part of the spectrum
remains approximately constant in flux and spectral shape and is in good
approximation independent of the $\gamma$-ray spectral variations, whereas the high-energy ($>1$~PeV) neutrino spectrum is variable. Its
relation to the photon flux is investigated below. 
\begin{table}
   \centering
   \caption{Pearson's correlation coefficient $r$ for the 1-50~PeV (0.1-1 PeV)
            neutrino flux ($F_\nu$) vs. the photon flux ($F_{\rm ph}$) in
            different energy bands.  The null hypothesis is that the true correlation between the fluxes is non-zero.}
  \begin{threeparttable}
  \begin{tabular}{cccc}  
  \hline
    & Energy band & $r(11)$\tnote{a}&  Remark  \\
  \hline
$F_\nu$ (1-50 PeV) &  &   \\
\hline
$F_{\rm ph}$  & $>200$~GeV & 0.97 &  S\tnote{b} \\
              & $0.1-300$~GeV& 0.94 & S \\
              & $15-50$~keV & 0.89 &  S \\
              & $2-10$~keV & 0.93 & S \\
\hline
$F_\nu$ (0.1-1 PeV) &  &   \\
\hline
$F_{\rm ph}$  & $>200$~GeV & -0.50 &  NS \\
              & $0.1-300$~GeV& -0.00&  NS  \\
              & $15-50$~keV & -0.43 &  NS \\
              & $2-10$~keV & -0.26 &  NS\\
\hline                                     
  \end{tabular}
   \tnote{a} The degrees of freedom (dof=N-2) for the Pearson's correlation
   significance test is given in the parenthesis. \\
   \tnote{b} For N=13, an observed value of $r$ larger than $\pm0.55$ is
   statistically significant (S) at a 5\% level for a non-directional hypothesis; otherwise, the correlation is non-significant (NS).
   \\
   \end{threeparttable}
\label{table2}
\end{table} 
Figure~\ref{fig9} shows the time evolution of the photon (black symbols/lines)
and neutrino (red symbols/lines)  fluxes in different energy bands as derived
by modeling the daily SEDs of \mrk.\ In particular, the daily fluxes are
shown as symbols, while continuous lines are the result of interpolation.
As a posterior check of our SED modeling, we verified that the relation  between the VHE $\gamma$-ray and X-ray (2-10~keV) fluxes is linear, in agreement
with the results reported in \cite{aleksic15}. The figure reveals signs of a  correlation between the
high-energy  neutrino flux and the photon fluxes in all energy bands. 
To quantify these findings, we performed a Pearson's correlation test (see
Table~\ref{table2}). The results show that the correlation  between  the
1-50~PeV neutrino flux and photon fluxes, in all energy bands, is statistically
significant for a non-directional hypothesis at a 5$\%$  level.  The strongest
correlation is found for the 1-50~PeV neutrino flux and the VHE ($>200$~GeV)
photon flux, with a Pearson's correlation coefficient $r=0.97$.  Furthermore,
the high-energy neutrino flux (in logarithmic units) can be adequately
described by a linear function of the logarithmic photon flux. For the 0.1-300~GeV flux, in particular, we find
\eqb
\log F_{\nu} = A\log F_{\gamma}+B,
\label{linear}
\eqe
where $F_{\gamma}$ is defined as the $\gamma$-ray flux in the 0.1-300~GeV
energy band, $A=1.59\pm 0.17$ and $B=5.25\pm1.64$. The relation between
$F_{\nu}$ and $F_{\gamma}$ is of particular interest for the estimation of the
cumulative neutrino event number  within the five years of full IceCube livetime (see
\S\ref{sec:longterm}).
\begin{figure}
    \centering
    \includegraphics[width=0.48\textwidth, height=0.35\textwidth]{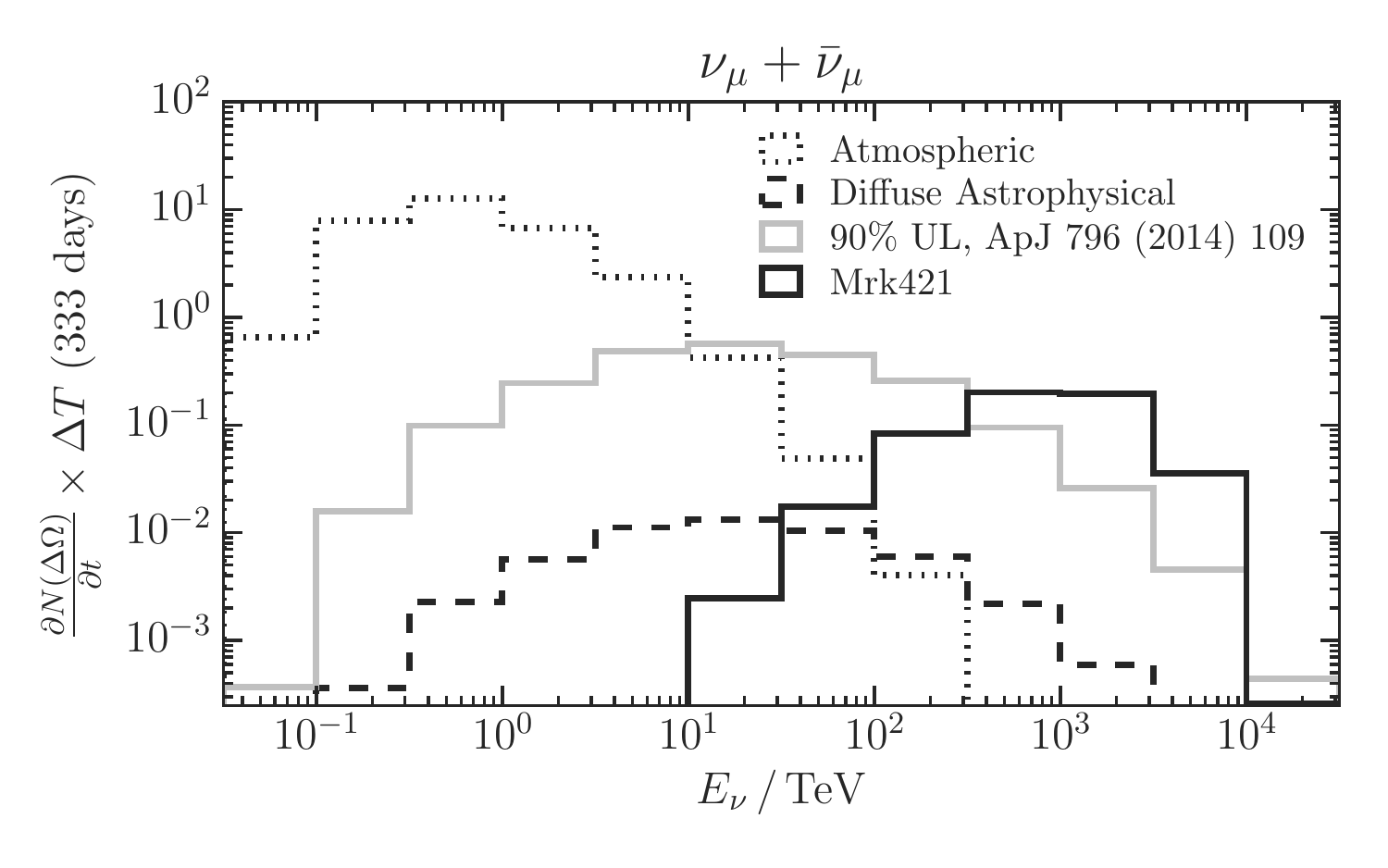}
    \caption{Differential (in energy) $\nu_\mu+\bar{\nu}_\mu$ event counts
             for 333~days of livetime calculated using
             eq.~(\ref{eq0}).  Three different components are shown: the
             atmospheric neutrinos produced in decays of charged
             $\pi^\pm/K^\pm$ (dotted line), the astrophysical  component
             observed by IceCube (dashed line) \cite{aartsen14, aartsen15},
             and the model-derived $\nu_\mu+\bar{\nu}_\mu$
             neutrino flux of \mrk \ for the quiescent period MJD~54850-54983
             (solid black line). The grey line shows the current IceCube 
             upper limit calculated using an unbroken $E^{-2}$ power-law
             spectrum \citep{aartsen14ps}. The observation window size
             corresponds to $\Delta\Omega=1^\circ$. Data are
             adopted from \citep{aartsen14ps}.}
    \label{fig:ev_rate}
\end{figure}
\begin{table*}
\centering
    \caption{Expected IceCube neutrino event rate for the
    $\nu_\mu+\bar{\nu}_\mu$ daily SEDs shown in Fig.~\ref{fig8} compared to the
    background event count rate. For the point spread function, a $90\%$
    angular resolution of 1\textdegree \, was assumed.  
    {All neutrino event rates are} in
    units of $\mathrm{yr}^{-1}$ assuming a good IceCube runtime of
    $\sim333\,\mathrm{days}$ per year, same as for the most recent point source
    data \cite{aartsen14ps}.}
    \begin{threeparttable}
    \begin{tabular}{ccccc}
        \hline
        & \multicolumn{2}{c}{\mrk\tnote{a}} & \multicolumn{2}{c}{Background\tnote{b}}  \\
        \hline
        $E_{\nu}$ (TeV) & 13-day flare  & quiescent &   atmospheric & diffuse \\
                        & (55265-55277) & (54850-54983) &  &         \\
        \hline                
        $0.1-100$       &    0.023          &       0.019       &  7.371 &  0.010 \\
        $100-10^3$      &    0.264          &       0.282	  &  $1.852\times10^{-3}$ & $2.203\times10^{-3}$ \\
        $10^3-5\times10^4$&  0.306	     &       0.288       &   $4.554\times10^{-6}$ &  $2.236\times10^{-4}$\\
        \hline
        \end{tabular}       
        \tnote{a} $90\%$ of the signal flux is expected to be within
                  $\Delta\Psi<1^\circ$.\\
        \tnote{b} Integrated over the bin-size $\Delta\Psi<1^\circ$.
     \end{threeparttable}
    \label{table3}
\end{table*}

Using the model-derived daily neutrino fluxes\footnote{The neutrino flux
produced at the source contains neutrinos of different flavors with an
approximate ratio $F_{\nu_{\rm e}}:F_{\nu_{\mu}} : F_{\nu_{\tau}} = 2 : 1 : 0$.
However, by the time they reach Earth their ratio will have changed to
$F_{\nu_{\rm e}}:F_{\nu_{\mu}} : F_{\nu_{\tau}} = 1 : 1 : 1$ due to neutrino
oscillations \citep{learned95}.} shown in Fig.~\ref{fig8}  and  after taking
into account the background neutrino fluxes, as described above, we calculate
the expected event rates $\dot{N}_\nu$ using eq.~(\ref{eq0}) at different
neutrino energy bins: 0.1-100~TeV, 0.1-1~PeV and 1-50~PeV.  The results for the
13-day flare are  summarized in Table~\ref{table3}. For comparison reasons, we
also included the expected differential event rate $\dot{N}_\nu^{q}$ for the
quiescent period MJD~54850-54983.  The hard neutrino spectra predicted by the
model for \mrk \ suggest that most of the signal neutrinos should be observed
at high energies, e.g. $\sim0.57$ evt/yr (events per year) above 100~TeV.  At
these energies, the neutrino background (atmospheric plus  astrophysical) is
negligible due to the soft energy spectrum and small observation window,
thus making a potential neutrino signal from \mrk \ a significant component. 

One should note, though, that at energies above 1~PeV Earth absorption starts to affect the
incident neutrino flux. This is illustrated in Fig.~\ref{fig:ev_rate} 
(see also Fig.~\ref{fig:IceCube_mrs}), where
the differential (in energy) $\nu_\mu+\bar{\nu}_\mu$ event counts  
for 333~days of livetime calculated using
eq.~(\ref{eq0}) for the quiescent state of \mrk,  are compared
against those from various backgrounds. The 90\% upper limit for \mrk \ as obtained by
IceCube \cite{aartsen14ps} is also plotted (grey histogram) for comparison reasons.
This is calculated assuming an unbroken $E^{-2}$ power-law neutrino
spectrum. Being much steeper than our model-predicted spectra, it 
yields less events above $\sim300$~TeV but predicts significantly
more neutrinos in the TeV - 300~TeV region, where IceCube shows the best
performance regarding point source searches (Fig.~$3$ in \cite{aartsen14ps}).

A comparison between the 13-day flare and  the
quiescent period reveals a net gain in the expected neutrino event rate of the
flare, at least for $E_\nu>1$~PeV.  This is, however, compensated by a relative
loss at energies $0.1-1$~PeV, thus leading to an approximately constant
neutrino event rate at energies $E_\nu>$100~TeV.  Although the $\gamma$-ray
activity of the source during the 13-day flare is high, e.g. the VHE
$\gamma$-ray flux varies by a factor of $\sim 4$ (see Fig.~\ref{fig9}) with a
peak flux reaching $\sim 2$~Crab~units \citep{aleksic15}, the neutrino event
rate of the flare in the background-suppressed regime is similar to the event rate of
the longer and non-flaring period.  We finally note that the biggest relative
gain is observed at lower energies ($E_\nu < 100$~TeV) where the high
atmospheric background reduces, however, the sensitivity. 

\begin{figure}
    \centering
    \includegraphics[width=0.48\textwidth]{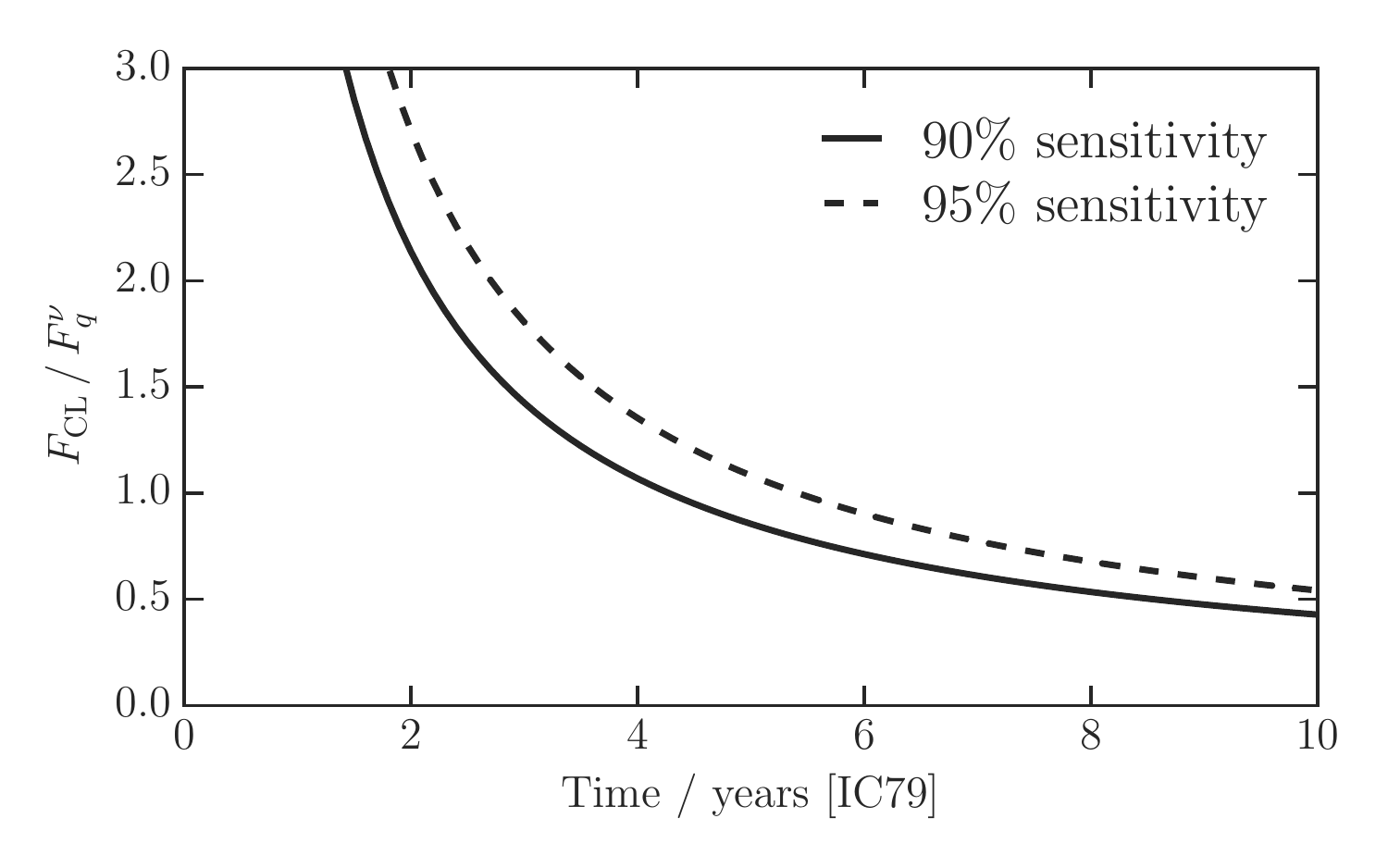}
    \caption{{Muon} neutrino flux $F_\nu$  from \mrk \ (in units of the quiescent neutrino flux
             $F_\nu^q$) as a function of the time needed for IceCube to observe neutrinos with
             energy $E_\nu>100\,\mathrm{TeV}$ at $90\%$ ($95\%$) confidence
             level. Time is measured in years with respect to the start
             date of IC79 ($\sim$MJD 55348).}
    \label{fig:time_sens}
\end{figure}

In the high-energy regime ($E_\nu>$100~TeV) a mean rate
$\dot{N}_\nu\sim0.57$ evt/yr is expected over a negligible background rate of 0.04 evt/yr.
Neglecting the small background, events originating from \mrk \ will be
detected at $90\%$ confidence, as soon as the expected total number of events
$N_\nu=\dot{N}_{\nu}\times T > 2.3$, where $\dot{N}_\nu$ is the neutrino event
rate and $T$ is the observation time. Figure~\ref{fig:time_sens} shows the flux
scaling needed for the models shown in Fig.~\ref{fig8} in order to observe
at least one event from \mrk \ at $90\%$ (solid line) and $95\%$ (dashed line) confidence level (CL). 
Given the IceCube event rate derived from the quiescent flux $F_\nu^q$, 
we find that IceCube should observe events above 100~TeV originating from \mrk
\ within 5 (6) years since the start of the 79-string IceCube (IC79) detector operation at  $90\%$ ($95\%$) CL.

\begin{figure*}
    \centering
    \includegraphics[width=0.9\textwidth, height=0.6\textwidth]{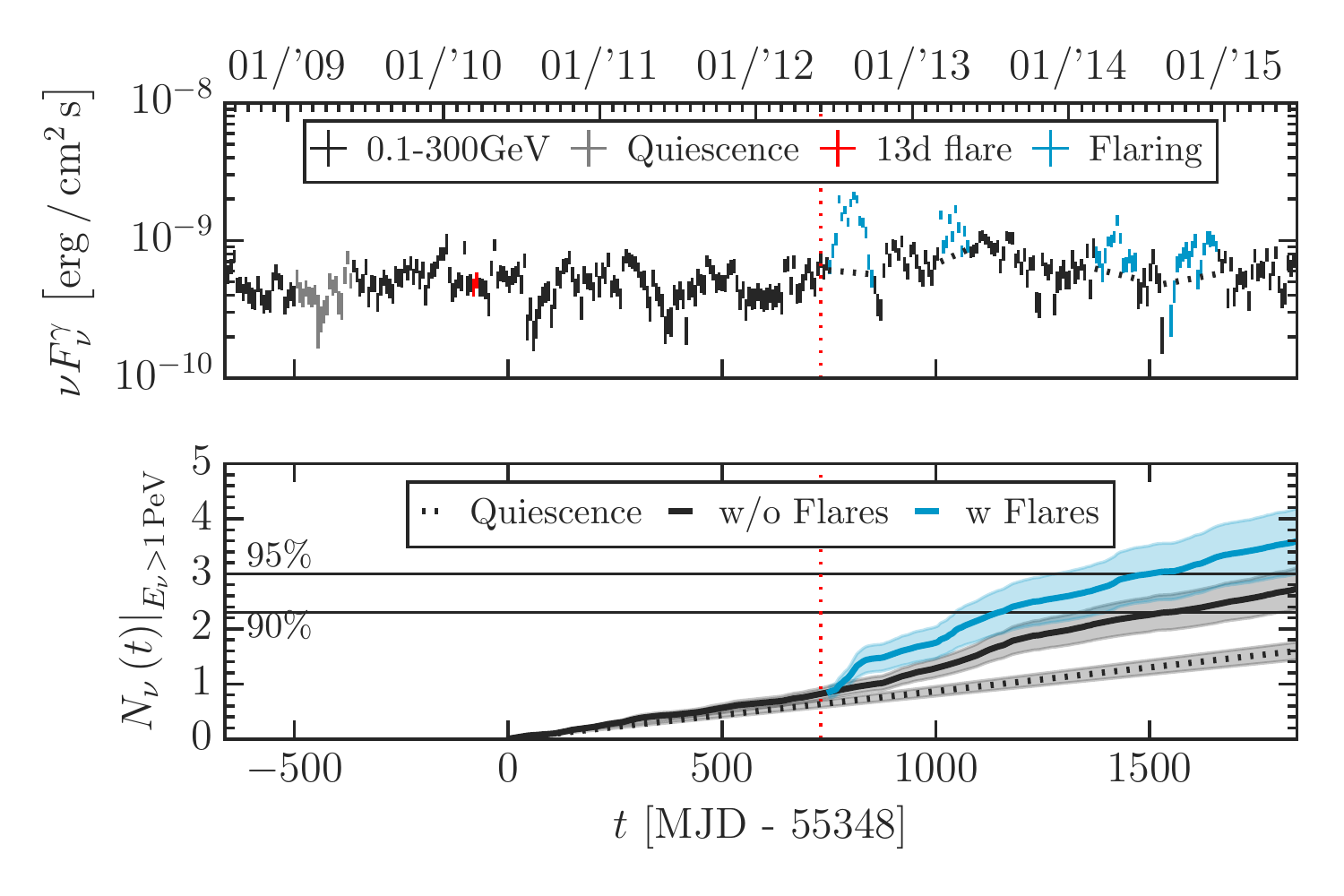}
    \caption{Top panel: Long-term, weekly binned  $\gamma$-ray light curve of
             \mrk \ at 0.1-300~GeV as observed with \fermi. \ The quiescent
             period and the 13-day flare are highlighted with grey and red symbols,
             respectively. At least four major flares (for the definition, see text) can be identified (light blue symbols).
              Bottom panel: The cumulative number of muon neutrino events above 1 PeV
             expected for IceCube within time $t$.
             The calculation is performed using the $\nu_\mu+\bar{\nu}_\mu$ flux estimated by the $\gamma$-ray light curve (top panel).
             The $\nu_\mu+\bar{\nu}_\mu$ flux is assumed to correlate with the $\gamma$-ray flux according to
             eq.~(\ref{linear}). The cumulative curves obtained with and without  the major flares included
             in the analysis are plotted  with thick light blue (`w flares') and black (`w/o flares')  lines, respectively.  The results of the  
             `quiescent analysis', where a constant $\gamma$-ray flux, 
             equal to that of the quiescent period (grey points in top panel) is assumed, 
             are also plotted for comparison (dotted line). {The latter is also the cumulative curve for neutrinos with energies 100 TeV - 1 PeV.}  In all cases, the $90\%$ uncertainties on the mean expected count are shown as shaded bands. These take
             into account the systematic uncertainties of the IceCube effective area, the statistical
             uncertainties of the $\gamma$-ray observations, and the error of the slope in the relation $\log F_\nu - \log F_\gamma$. Horizontal lines indicate the  threshold for the observation of one or more neutrinos
             at  $90\%$ or $95\%$ CL. The latest published results of IceCube \cite{aartsen14ps} included data until MJD 56063, which is marked by the vertical red-dotted line.
             }
    \label{fig:long-term}
\end{figure*}

\section{The long-term $\gamma$-ray activity}
\label{sec:longterm}
During the 13-day flaring period of 2010 the neutrino flux above 1~PeV was
found to correlate with the photon flux in all energy bands under
consideration. Assuming that the correlation is present in longer time periods
as well, we may apply the derived linear relation between the logarithmic
neutrino and photon fluxes to the long-term light curve at a specific energy
band, in order to calculate the expected number of muon neutrino events at
$E_{\nu}\ge1$~PeV within the five years of full IceCube detector
livetime.\footnote{ Here, the full IceCube livetime is defined with respect to
the start of operation with 79 strings in May/June 2010 (IC79), i.e.  before
the completion of the detector one year later with 86 strings.}

Although the strongest correlation was derived for the VHE $\gamma$-ray and
high-energy neutrino fluxes, here we choose to perform our analysis on the
long-term \fermi \ (0.1-300~GeV) light curve of \mrk. The reason for doing so
is that the {\sl Fermi}-LAT light curve (MJD 54686.15-57192.15) covers the
period of the complete IceCube operation, whereas the available long-term light
curves in VHE $\gamma$-rays extend at most up to 2009 (for details, see
\cite{fraijamarinelli15}).

Figure \ref{fig:long-term} (top panel) shows the long-term (2506 d), weekly
binned $\gamma$-ray light curve as observed with \fermi. \ The quiescent period
of 2009 and the 13-day flare of 2010 are highlighted with grey and red symbols,
respectively. Starting with a major $\gamma$-ray flare detected by \fermi \ in
summer of 2012  \citep{dAmmando2012}, \mrk \ entered a prolonged, still
on-going, high-state period.  During this period, at least four flares
with $\gamma$-ray fluxes up to 3-10 times higher than that of the 2010 flare can
be identified (light blue symbols). Henceforth, these will be referred to as `major' flares.
Inspection of the \fermi \ light curve alone would question the definition of the period MJD~55265-55277 as a flare.
Although no significant variability was detected in the \fermi \ energy band (see also Fig.~\ref{fig9}),
the X-ray and VHE $\gamma$-ray variability was remarkable \citep{aleksic15}
(see also Figs.~\ref{fig1} and \ref{fig2}).

Each of the major flares was fitted with a Lorentzian function and its characteristic duration was defined
as twice that corresponding to its FWHM.  The derived times of the peak fluxes and the respective flare durations are listed below:
\begin{itemize}
    \item Flare 1a: MJD~$56122.9\pm26.9$
    \item Flare 1b: MJD $56155.7\pm43.9$
    \item Flare 2: MJD $56393.09\pm34.04$ 
    \item Flare 3: MJD $56767.2\pm47.6$ 
    \item Flare 4a:  MJD $56942.3\pm62.3$ 
    \item Flare 4b:  MJD $56992.5\pm79.9$
\end{itemize}
We note that flares 1a and 1b partially overlap and can be considered as one
major double-peaked flare (see also \citep{hovatta2015}). The same applies to
the latest flare. Its long duration as determined by a single Lorentzian fit to
the data, suggests that this consists of, at least, two overlapping flares,
here noted as Flares 4a and 4b.

The expected number of muon neutrino events can be then calculated as follows.
The average $\gamma$-ray flux of the quiescent period MJD 54850-54983 is defined as
\eqb
    F_\gamma^q \equiv \frac{1}{T_q}\int_{T_q} dt F_\gamma(t),
    \label{Fgamma_q}
\eqe
where $T_q=133$~d and
$F^q_\gamma=4.199^{+0.175}_{-0.165}\times 10^{-10}$~erg~cm$^{-2}$~s$^{-1}$ (for the error
calculation, see \ref{appenC}).  The model-predicted $\nu_\mu+\bar{\nu}_\mu$ event rate above 1 PeV
for the quiescent period of \mrk \ is $\dot{N_{\nu}}^q=0.288$ evt/yr (see
also Table~\ref{table3}). If $\dot{N}_{\nu}$ is the neutrino event rate for the
period $T$, then $\dot{N}_{\nu} = \dot{N}_{\nu}^{q} F_{\nu}/F_{\nu}^{q}$, where
$F_{\nu}^{q}$, $F_{\nu}$ are the average neutrino fluxes for the periods $T_q$
and $T$, respectively, and are defined similarly to eq.~(\ref{Fgamma_q}).
According to the linear correlation derived in the previous section, the
$\gamma$-ray and neutrino fluxes are related as $\log F_\nu=A\,\log
F_\gamma+B$. Thus, using the long-term \fermi \ light curve, we may estimate
the number of events expected in time period $T$ as  
\eqb
    N_\nu \equiv \dot{N}_\nu T = \frac{\dot{N}_{\nu}^q}{F_\nu^q} \int_T dt\,F_\nu\left(t\right)
         =\dot{N}_{\nu}^{q}\int_{T} dt\,\left(\frac{F_\gamma\left(t\right)}{F_\gamma^q}\right)^A.
    \label{eq:Nnu1}
\eqe
As the $\gamma$-ray light curve of \mrk \ is weekly binned, the integral of
eq.~(\ref{eq:Nnu1}) can be approximated by a sum with bin-width $\Delta t=7$~d:
\eqb
    N_\nu=\frac{\dot{N}_{\nu}^{q}}{F_\nu^{q}} \int_T dt\,F_\nu\left(t\right)
         =\frac{\dot{N}_{\nu}^{q}\Delta t}{\left(F_\gamma^{q}\right)^A}\sum_i \left(F_{\gamma,i}\right)^A.
    \label{eq:Nnu2}
\eqe
The logarithmic event count ${n}_{\nu}=\log {N}_\nu$
then yields
\eqb
    n_{\nu} =\log\left(\Delta t \,\dot{N}_\nu^{q}\right)
            + \log \sum_i \left(F_{\gamma,i}\right)^A - A \log F_\gamma^{q},
    \label{eq:Nnu3}
\eqe
and the respective error is given by
\eqb
\sigma^2_{n_{\nu}} = f^2_{\dot{N}_{\nu}^{q}} +  f^2_{F_{\gamma,i}} + f^2_{F_{\gamma}^{\rm q}} + f^2_{\rm A},
%
\label{eq:error1}
\eqe
where the various contributions to the total uncertainty are
\eqb
\label{eq:err-Nq}
f_{\dot{N}_{\nu}^{q}} & = & \frac{\sigma_{\dot{N}_{\nu}^{q}}}{\ln 10\, \dot{N}_{\nu}^{q}} \\
\label{eq:err-Fg}
f_{F_{\gamma,i}} & = & \frac{\sqrt{\sum_i \left(\sigma_{F_{\gamma,i}} A F_{\gamma,i}^{A-1}\right)^2}}{\ln 10\, \sum_i F_{\gamma,i}^A} \\
\label{eq:err-Fq}
f_{F_{\gamma}^{\rm q}} & = & \frac{A\sigma_{F_{\gamma}^{q}}}{\ln 10 \, F_{\gamma}^{q}} \\
\label{eq:err-A}
f_{\rm A} & = & \frac{\sigma_A\sum_i F_{\gamma,i}^A \ln\frac{F_{\gamma,i}}{F_{\gamma}^{\rm q}}}{\ln10\,\sum_i F_{\gamma,i}^A}.
\eqe
In the above, $\sigma_{\dot{N}_\nu^{q}}$ is the uncertainty of the muon
neutrino event rate in quiescence, which is dominated by systematic effects of
the IceCube detector, and is accounted for with $10\%$ relative uncertainty
\citep{aartsen14ps}. Furthermore, $\sigma_{F_{\gamma,i}}$ is the statistical
error of the $\gamma$-ray flux measurements and $\sigma_{F_{\gamma}^q}$ is the
uncertainty of the $\gamma$-ray flux in quiescence (for the derivation, see
\ref{appenC}). Finally, $\sigma_A$ is the $1\sigma$ error of $A$ as obtained
from fitting the modeled neutrino and 0.1-300~GeV $\gamma$-ray fluxes (see
eq.~(\ref{linear})).

Using eqs.~(\ref{eq:Nnu2}) , (\ref{eq:Nnu3}) and (\ref{eq:error1}), we
calculated $N_\nu$ in three different cases described below. First, we
performed the analysis using the full \fermi \ light curve (`w flares'
analysis). Then, to exemplify the net effect of the major flares on the
expected number of muon neutrino events, we performed the analysis after
excluding the major flares  (`w/o flares' analysis). However, in order to include the
respective exposure time in this analysis, we interpolated the $\gamma$-ray light curve using
mean fluxes  in the time just before and after the end of each major
$\gamma$-ray flare (see black dashed lines in top panel of
Fig.~\ref{fig:long-term}).  Finally, we considered the extreme case of a
non-variable $\gamma$-ray light curve with flux equal to that of the 2009
quiescent period (`quiescent' analysis). 
\begin{figure}
    \centering
    \includegraphics[width=0.48\textwidth]{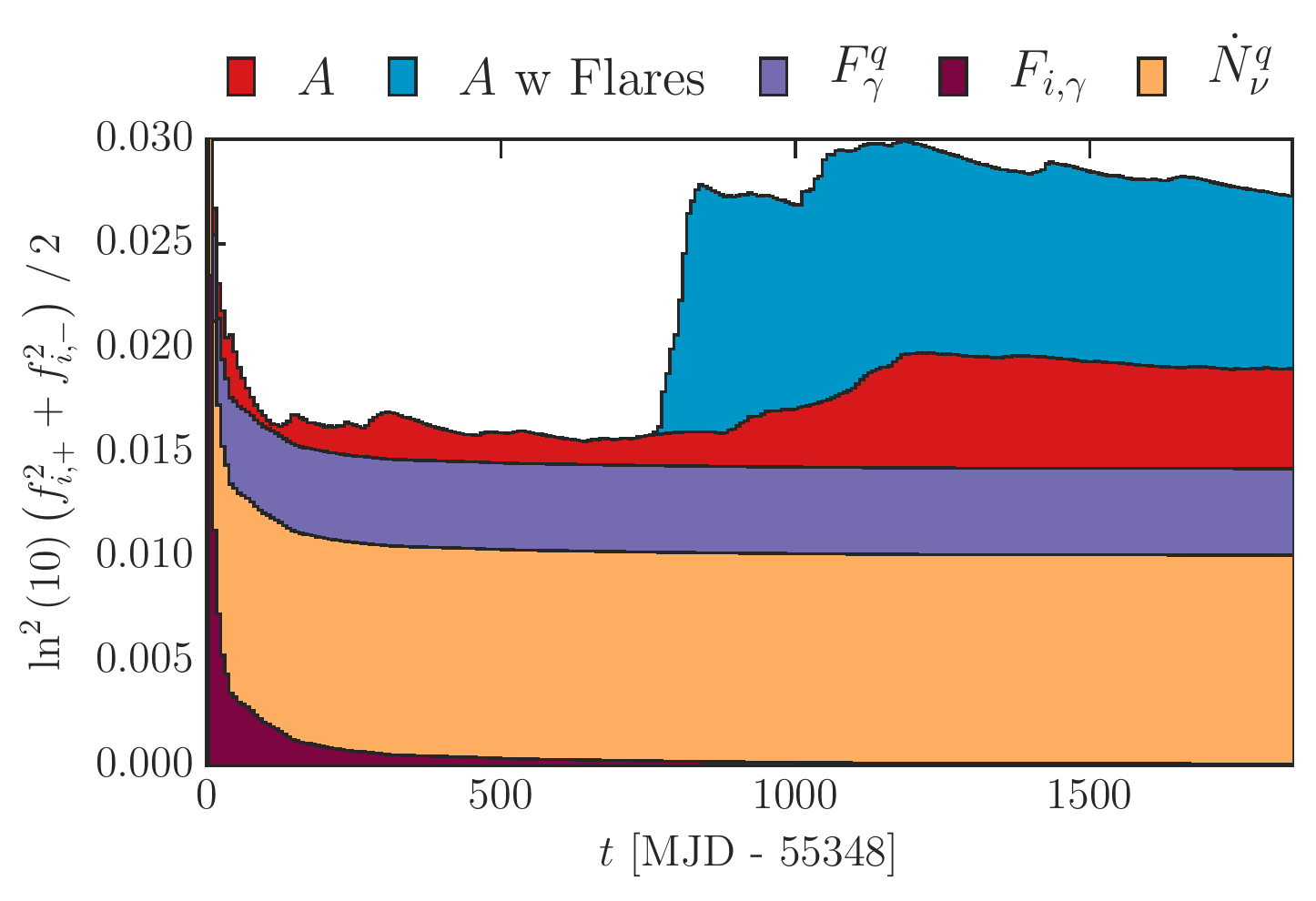}
    \caption{
             Stacked contributions of the various sources of uncertainty $f_i^2$
             defined in eqs.~(\ref{eq:err-Nq})-(\ref{eq:err-A}) that  add up to the
             total uncertainty given by eq.~(\ref{eq:error1}). In addition, the
             uncertainty on the slope $A$ of the $\log F_{\nu}-\log F_{\gamma}$ relation is shown for the two scenarios, i.e. with
             and without major flares in blue and red, respectively. For all other sources of uncertainty, the differences between the two scenarios are negligible. 
             }
    \label{fig:error}
\end{figure}

The cumulative event count $N_\nu$ expected for IceCube from \mrk \ is presented in the bottom panel of Fig.~\ref{fig:long-term}, where 
the results of the different analyses are plotted with different types of lines (for details, see figure caption). 
The total uncertainty given by eq.~(\ref{eq:error1}) is shown, in each case, as a shaded
band around the mean event count. The various sources of uncertainty that contribute to the total one are presented
in Fig.~\ref{fig:error}. The major contributions to  $\sigma^2_{n_{\nu}}$ originate from 
the systematic uncertainty of $A_{\rm eff}$ (yellow pale color) and from the uncertainty on 
the $\log F_\nu-\log F_\gamma$ relation. Although the latter is sub-dominant in the analysis `w/o flares' (red color) compared to the systematic uncertainty, it becomes comparable to it when the major flares are included  (blue color). 
The statistical uncertainty of the $\gamma$-ray flux  measurements is similar in both analyses, while it becomes negligible as the total observing time in $\gamma$-rays increases.

From Fig.~\ref{fig:long-term} (bottom panel), it becomes evident that the inclusion
of  major $\gamma$-ray flares greatly increases the event number, as long as
the derived correlation of the $\gamma$-ray flux with the $>1$~PeV neutrino
flux still holds.  To better quantify these results, we present in
Tables~\ref{table4} and \ref{table5} the expected $\nu_\mu+\bar{\nu}_\mu$ event
numbers for each  one of the  IceCube operation seasons and major flares, respectively\footnote{The respective numbers of $\nu_\mu+\bar{\nu}_\mu$ events with 100 TeV$<E_\nu < 1$~PeV can be easily obtained from the values listed in Table~\ref{table3} and the duration of each operation season.}.  
 We find that the event rate during Flares 1a and 1b exceeds
the event rate for the same year (06/2012-05/2013) by a factor $3.0\pm0.9$. 
{Although the probability to observe at least one muon neutrino event during Flares 1a and 1b is $46\%$ (see Table~\ref{table5}), a restricted neutrino search over the period of the 2012 major flare would not guarantee a neutrino detection. In contrast, $1.46 \pm 0.32$ $\nu_\mu+\bar{\nu}_\mu$ events are expected within the period covered by all four major flares (385 days in total). In this regard, stacked analyses of major flares  from Mrk 421 would be more beneficial for neutrino searches.}

These results, albeit model-dependent, demonstrate that major (in duration and flux) flares from blazars like \mrk \ could serve as favorable periods for time-dependent neutrino searches. Interestingly, time-dependent searches of IceCube  \cite{aartsen15_time} exist until MJD~56063 (red dotted line in
Fig.~\ref{fig:long-term}), that is just before the major flare of \mrk \ in 2012.

Excluding the flares as explained previously, the expected number of
$\nu_\mu+\bar{\nu}_\mu$ events up to the date where the most recent IceCube
data are available (05/2015),  is found to be $2.73\pm0.38$ (see
Table~\ref{table4}). This exceeds the $90\%$ threshold value for detection within the
uncertainties. The respective number in the analysis with the major flares
included increases to $3.59\pm0.60$, which
excludes a non-detection of neutrinos by more than $95\%$.
By utilizing the neutrino-photon flux correlation, we
found a significant increase in the expected neutrino rate compared to that
obtained from the quiescent state alone. In particular, the prediction of the
quiescent state would yield $\sim1.6$ events in the IceCube livetime, thus
underestimating  the neutrino event rate after June 2012. Since then, \mrk \
entered a high $\gamma$-ray flux period that is still on-going (see top panel
in Fig.~\ref{fig:long-term}).  We remark that the estimate of $1.6$ events
applies also to neutrinos with energies in the range 100~TeV-1~PeV, since we
found no correlation between the photon and sub-PeV neutrino fluxes in our
modeling of the 13~day flare (see Table~\ref{table2}). In this regard, the
estimate derived from the quiescent state is the most robust.

\begin{table}
  \centering
  \caption{Number of high-energy $\nu_\mu+\bar{\nu}_\mu$ events ($E_\nu>1$~PeV)
           expected for IceCube in various seasons of operation (each with
           duration T in days).  The results are obtained using the \fermi\
           $\gamma$-ray light curve in Fig.~\ref{fig:long-term} (top panel) and
           the connection to the high-energy neutrino flux through
           eq.~(\ref{eq:Nnu1}). The values for each season are obtained after
           replacing the major $\gamma$-ray flares (Table~\ref{table5}) with a
           non-variable emission, whose flux was determined through
           interpolation of the $\gamma$-ray light curve just before the start
           and after the end of each major flare (for details, see text).  The
           total number of events without (with) the major flares included are
           also presented. For each entry, the probability $P_{N_\nu\geq1}$ of
           observing one or more neutrinos is quoted.}
  \begin{threeparttable}
    \begin{tabular}{cccc}
        \hline
        Season & T (days) & $\nu_{\mu}+\bar{\nu}_{\mu}$ & $P_{N_\nu\geq1} (\%)$\tnote{$\dagger$}\\
         \hline        
        06/2010-05/2011 &  364 & $0.43\pm0.06$ & $34\pm4$ \\
        06/2011-05/2012 &  364 & $0.38\pm0.05$ & $32\pm3$ \\
        06/2012-05/2013 &  371 & $0.71\pm 0.11$ & $51\pm5$ \\
        06/2013-05/2014 &  364 & $0.70\pm 0.11$ & $50\pm5$ \\
        06/2014-05/2015 &  350 & $0.47\pm 0.06$ & $38\pm4$\\
        \hline
        $\sum$ w/o Flares & 1834\tnote{a} & $2.73\pm0.38$ & $94\pm2$ \\     
        $\sum$ w Flares  & 1834 &  $3.59\pm0.60$ & $97\pm2$  \\
        \hline
    \end{tabular}
    \tnote{$\dagger$} Using Poisson statistics, $P\left(N_{\nu}\geq1\right)=1-\mathrm{e}^{-\lambda}$
    for a Poisson distribution with mean $\lambda$.\\
    \tnote{a} On top of the quoted years, three weeks of additional \fermi \
    data are available after 05/2015.
  \end{threeparttable}
  \label{table4}
\end{table}
\section{Discussion}
\label{discussion}
\begin{table}
  \centering
  \caption{Same as Table~\ref{table4} but for the four flares that were identified
           in this analysis.}
    \begin{tabular}{cccc}
        \hline
        No. & T (days) & $\nu_{\mu}+\bar{\nu}_{\mu}$ & $P_{N_\nu\geq1}(\%)$\\
        \hline
        Flares 1a+1b     &  105  & $0.61\pm 0.16$ & $46\pm8$ \\
        Flare 2         &   70  & $0.32\pm 0.07$ & $27\pm5$ \\ 
        Flare 3          &   98 & $0.26\pm 0.05$ & $23\pm4$ \\ 
        Flares 4a+4b        & 112  & $0.26\pm0.05$ & $23\pm4$ \\
        \hline
        $\sum$ Flares  & 385  & $1.46\pm0.32$ & $77\pm7$  \\
        \hline
    \end{tabular}
  \label{table5}
\end{table}
By modeling the daily SEDs of \mrk \ we were able to derive the daily $\nu_\mu+\bar{\nu}_\mu$ neutrino fluxes
and compare them against those obtained for the longer, albeit quiescent, period of 2009. Although one could naively argue
that the neutrino event rate would be higher during the 13-day flare, we showed explicitly that the mean event rate above 100~TeV 
{is $\sim0.57$~evt/yr ($\sim$0.26 evt/yr for 100 TeV$<E_\nu<$1 PeV and $\sim$0.31~evt/yr for $E_\nu > 1$~PeV). This is } comparable to that expected for a four-month period of lower $\gamma$-ray and X-ray fluxes. Due to the short duration of the flare, the expected number of muon neutrino events is $\sim 0.02$, i.e. insufficient to explain a fiducial neutrino detection from the direction of \mrk \ in the time-window of a flare with similar characteristics as the one studied here. Interestingly, \cite{reimeretal05} reached similar conclusions within a different leptohadronic model for explaining the orphan VHE $\gamma$-ray flare of blazar 1ES 1959+650.

Inspection of the \fermi \ light curve (top panel in Fig.~\ref{fig:long-term})
clearly shows a transition of the source to a period of increased $\gamma$-ray
flux (i.e., high state) that was initiated in 2012 by a major flare, and is
still on-going. Within this period we identified four major flares, in total,
with peak fluxes $\sim 3-10$ times higher than the one modeled here. Whether
the correlation between the neutrino and $\gamma$-ray fluxes we derived in
\S\ref{subsec:neutrino} still holds during these extreme flares cannot be
safely answered without detailed modeling of the respective SEDs. As the SED is
composed of many emission components  (see Fig.~\ref{fig1}), which are moreover
dependent on each other, it is not trivial to predict the neutrino-$\gamma$-ray
flux correlation during flares without having knowledge of the variability at
lower energy bands; this was the motivation of this study in  the first place.
Consider for example a scenario where the major $\gamma$-ray flare is not
accompanied by a respective increase of the X-ray flux. An increase of the
proton injection luminosity, which would also lead to a higher neutrino flux,
could not explain this fiducial flare, since the Bethe-Heitler component would
also be enhanced, therefore violating the fiducial observations.
More than one model parameter should be changed and,
depending on their combination, the derived neutrino flux would also differ.
Because of the wide range of possibilities, in the present study we simply
assume that the relation given by eq.~(\ref{linear}) is valid over the 6.9~yr
period of \fermi \ observations, while the major flare MW modeling and its
implications for the neutrino flux will be the subject of a subsequent paper.

Nevertheless, in order to assess the net effect of the major flares on the
predicted $\nu_\mu + \bar{\nu}_\mu$ event number we performed an additional
analysis (`w/o flares'), where these were not taken into account. More
precisely, in order to include the respective exposure times in our analysis,
we replaced the high $\gamma$-ray fluxes of the major flares with values
obtained from the interpolation of the $\gamma$-ray light curve just before the
start and after the end of each major flare (see dashed lines in the top panel
of Fig.~\ref{fig:long-term}).  We showed that
$1.46\pm0.32$ $\nu_\mu+\bar{\nu}_\mu$ events are expected within a period of
$385$~days due to the major flares alone. Thus, their presence increases the
neutrino event rate within the IceCube livetime by $30\%$ (see
Tables~\ref{table4} and \ref{table5}).  Furthermore, the neutrino rate without
(with) the major flares, as estimated by the long-term $\gamma$-ray light
curve,  is $73\%$ ($127\%$) higher than that expected by simply extrapolating
the neutrino flux in the quiescent state (see Table~\ref{table2}). In brief,
the predictions for the cumulative event count above 1 PeV are significantly
affected by the major flares under the assumption of a neutrino-$\gamma$-ray
flux relation given by eq.~(\ref{linear}). The values derived from the
`quiescent' analysis ($1.60\pm 0.16$ events in 1834~days) apply, however,
directly to the 100~TeV-1~PeV neutrino event counts, since we found no
significant correlation between the sub-PeV and photon fluxes (in any energy
band). Meanwhile, they constitute a robust lower bound for the predicted
cumulative events above 1~PeV.

{Based on four years of data {searches using through-going muons}, IceCube reported an overfluctuation of events
at the position of \mrk \ \cite{aartsen14ps}. The best-fit to the data yielded
$n_{\rm s}=3.8$ signal events over the full energy range for an unbroken
power-law spectrum $dN_{\nu}/dE_{\nu} \propto E_\nu^{-\gamma}$ with
$\gamma\sim1.9$, while 22.4 background events were expected in a circle
1\textdegree \ around the search coordinates. 
This result deviates from the atmospheric background expectation of a soft
spectrum with $\gamma\sim3.7$, but is still consistent with a pure background
expectation ($p\approx26\%$, where $p$ is the pre-trial probability). The
$90\%$~upper limit on the flux normalization of an unbroken $E^{-2}$ flux was
set to $\Phi^{90\%}_{\nu_\mu+\bar{\nu}_\mu}
=2.1\times10^{-12}\,\mathrm{TeV^{-1} \, cm^{-2} \,s^{-1}}$.  As
Fig.~\ref{fig:ev_rate} demonstrates, such a soft spectrum yields events mostly
at the TeV energy range, whereas the model adopted in this study predicts a
spectrum peaking at the PeV energy range. Two out of the four years used in
\cite{aartsen14ps, aartsen15_time} are in full detector configuration, thus
coinciding with our calculations for the long-term light curve of \mrk \ (see
\S\ref{sec:longterm}). Within the overlapping period of 728~days,
$0.81\pm0.08$ $\nu_\mu+\bar{\nu}_\mu$ events are expected above PeV
energies and $0.62\pm0.06$ for energies between 100~TeV and 1~PeV (see
Table~\ref{table4}). These results do not contradict the observation of IceCube that
the overfluctuation is consistent with pure background. A fit to more recent IceCube data that include two additional years \cite{icrc2015ps}
increased the number of signal events to $n_{\rm s}=5.5$. Yet, this is still
consistent with the background expectation.

Within the period {of five years}, and according to our estimations, IceCube is expected to detect
high-energy ($E_\nu>1$~PeV) neutrinos from \mrk \ at $90\%$ confidence level. Hence, with additional data, IceCube's sensitivity will surpass our model
predictions, thus testing scenarios of cosmic-ray acceleration at the PeV 
energy regime. Even a non-detection of neutrinos would be of great importance, {though;  this can}
place constraints on the contribution of the hadronic component to
the high-energy emission from \mrk, {as we illustrate below. 
Given a non-detection in $X$ years, the most robust constraint on the hadronic contribution of \mrk \  can be derived from the quiescent scenario (see  e.g. Fig.~\ref{fig:time_sens}).  The rest of our predictions (see Tables~\ref{table4} and \ref{table5}) are based on the assumption of a correlation $F_\nu ~\propto F_\gamma^{A}$, whose long-term validity may be questionable. In order to be able to constrain the hadronic component in flaring periods, one should first test the validity of the correlation in different epochs of flaring activity through SED modeling. A smaller number of neutrinos implied by a non-detection in $X$ years could be  obtained either by an absent correlation or a lower proton luminosity in the blob. Hence, in the following, we focus on the quiescent scenario where the constant $\nu_{\mu}+\bar{\nu}_\mu$ flux is related to the proton luminosity as $F_\nu = (1/4) f  L_{\rm p}/ 4 \pi  d_{\rm L}^2$, where we assumed that 50\% of the interactions lead to  $\pi^0$ production and $L_{\rm p}$ 
is the proton luminosity in the blob as measured in the observer's frame. Moreover, $f$ is the pion production efficiency which may be written as  (see e.g. \citep{petromast15})
\eqb
f\approx 2.2 \times 10^{-3}\, L_{\rm s, 45} R^{-1}_{15} \nu^{-1}_{\rm s, 16} \delta_1^{-3},
\eqe
where $L_{\rm s}$ is the apparent bolometric luminosity of the low-energy hump of the SED and $\nu_{\rm s}$  is the respective peak 
frequency. Here, the notation $q_{\rm X}\equiv q/10^{\rm X}$ has been introduced. For parameters relevant to the quiescent period $f\approx 10^{-5}$, in agreement with previous studies on BL Lac neutrino emission (e.g. \citep{atoyandermer01, murase14, petroetal15}). A non-detection of muon neutrinos above 100 TeV in $X$ years translates into $F_{\nu, \rm X} = \zeta_{\rm X} \, F_{\nu}^{(\rm q)}$, where $F_{\nu}^{(\rm q)} \simeq 2.4 \times 10^{-10}$ erg cm$^{-2}$ s$^{-1}$ and 
$\zeta_X \le 1$ can be read from the sensitivity curves shown in Fig.~\ref{fig:time_sens}.  Combination of the above leads to 
\eqb
L_{\rm p, X} = 1740\, \zeta_{\rm X} \, F_{\nu, \mu}^{(\rm q)} 4\pi d_{\rm L}^2 L_{\rm s, 45}^{-1} R_{15} \nu_{\rm s, 16} \delta_1^3 \, {\rm erg \, s}^{-1}.
\eqe 
Our results are summarized in Table~\ref{table6} for two CL values.}
\begin{table}
  \centering
  \caption{Upper limits on the proton luminosity in the blob as derived from a 
  non-detection (at 90\% and 95\% CL) of muon neutrinos ($>100$ TeV) from \mrk \ in $X$ years. }
    \begin{tabular}{ccccc}
        \hline
        $X$ (yr) & \multicolumn{2}{c}{$\zeta_X$} & \multicolumn{2}{c}{$L_{\rm p, X}$ (erg/s)}\\
                & 90\% & 95\% & 90\% & 95 \% \\  
         \hline        
        6  & 0.71 & 0.9 &   $6.2\times 10^{47}$ & $7.8\times10^{47}$ \\
	8 & 0.53  & 0.68 &   $4.6\times10^{47}$& $5.9\times10^{47}$\\
        10& 0.43 & 0.54  &   $3.7\times10^{47}$& $4.7\times10^{47}$\\
	20 & 0.21 & 0.27 & $1.8\times 10^{47}$& $2.3\times10^{47}$\\
       \hline
    \end{tabular}
  \label{table6}
\end{table}


{
All the estimates we have presented so far are based on the up-going muon
sample, as this is the most relevant for point source searches, especially,
those located in the northern sky (for details, see \S\ref{sec:eventrate}). The
HESE sample, on the other hand, consists of a small, yet high purity,
statistical sample of astrophysical all-flavors neutrinos due to the veto
imposed on atmospheric events. Its high purity comes, though, at the cost of a
largely reduced effective area (see e.g. Fig.~\ref{fig:IceCube_effA}). Thus, a
meaningful comparison between the expected number of muon neutrinos listed in
Tables~\ref{table2}-\ref{table4} and depicted in Fig.~\ref{fig:long-term} with
those obtained from the five-year analysis of the HESE sample
\cite{KopperICRC2015} would require an appropriate scaling of the rates by a
factor of $\sim 50$, which is due to the difference in effective areas as shown
in Fig.~\ref{fig:IceCube_effA}. For the most optimistic scenario 
considered here (`w flares'), the expected number of $\nu_\mu+\bar{\nu}_\mu$ events in 5
years with the HESE analysis would read $\sim 3.59/50\simeq 0.07$ or $\sim 0.2$
all-flavor events. These values are still consistent with the results reported
in \cite{KopperICRC2015}.
}

It is also worth noting that our hypothesis of the PeV
neutrino-$\gamma$-ray correlation during major flares can be further tested with
specific, time-optimized analyses similar to those presented in
\cite{aartsen15_time}. {Nevertheless, multiple flares or long-lasting
flaring periods are needed to accumulate enough exposure for a neutrino detection.}
These can make use of the most recent IceCube data,
since enhanced neutrino event rates are expected, based on our hypothesis, in
the recent years of full IceCube detector operation. Furthermore, the long-term
\fermi \ light curve of \mrk \ implies that the quiescent state of 2009 may not be
the most characteristic state of activity, since the blazar entered a
long-lasting period of increased $\gamma$-ray flux since the summer of 2012. Consequently, the observation of
\mrk \ in search of high-energy neutrinos might be more efficient, when focused
on periods where higher neutrino emission is expected.

There are two additional mechanisms of neutrino production implied by 
our model, which were not shown in detail because their contribution to 
the total spectrum is minimal. The first is neutron decay. High-energy neutrons, a by-product of photomeson
interactions (e.g., $p+\gamma \rightarrow \pi^{+} + n$), are free to escape the emission region, thus providing an effective means of cosmic-ray escape \citep{kirkmast89,giovanonikazanas90,atoyandermer03}, while producing at the same time $\bar{\nu}_{\rm e}$.  As shown 
in DPM14, those neutrinos are lower in both energy and flux than the ones resulting from meson decays, by about two orders of magnitude. 
The second mechanism results from the $\beta$-decay produced protons 
propagating in the interstellar and intergalactic medium as cosmic rays, 
and interacting with ambient radiation fields \citep{stecker68}, i.e. the extragalactic background light (EBL). In the present discussion we neglect any contribution to the cosmic-ray flux from direct proton escape \citep[e.g.][]{baerwald13}. The highest-energy escaping
protons considered in our model have energy $E_{\rm p, \max}^{\rm esc} \sim 1.6 \times 10^8 \, {\rm GeV} \, \Gamma_{1.3}$, 
where $\Gamma \sim \doppler$. These protons are energetic enough
to pion-produce on photons with energy $E_0 \gtrsim 0.9 \, \rm{eV} \, \Gamma_{1.3}^{-1}$. The present ($z=0$) EBL energy density at $\sim 1$~eV is $u_0 \simeq 4\times 10^{-15}$~erg cm$^{-3}$ \citep[e.g.][]{steckerscully06,finke10,kneiske10}. The photopion energy loss rate for protons with $E_{\rm p, \max}^{\rm esc}$ can be then estimated as $t^{-1}_{\rm p\pi, EBL} \approx \kappa_{\rm p\pi} \sigma_{\rm p\pi} c n_0 E_0$, where $n_0 = u_0/E_0^2$, and $\kappa_{\rm p}=0.2$ and $\sigma_{\rm p\pi} \simeq 5\times 10^{-28}$cm$^2$ are the inelasticity and cross section, respectively, at the $\Delta(1232)$ resonance \citep{beringer12}. The pion production efficiency on the EBL photons can be then estimated as
$f_{\rm EBL} \equiv  t_{\rm cr}/t_{\rm p\pi, EBL} \approx 10^{-4}$, where we conservatively used $t_{\rm cr}=d_{\rm L}/c$, assuming rectilinear proton propagation. The propagation of protons at these energies may be diffusive \citep[e.g.][]{Lemoine2005}, thus increasing the residence time of protons in the ISM, while isotropizing both the cosmic-ray and accompanying  neutrino fluxes. Regardless, the net effect would be a decrease of the observed neutrino flux. 
The efficiency $f$ of pion production in the emission region of the  blazar jet can be calculated in a similar way.  By approximating the 
low-energy hump of the SED as a monochromatic photon field with characteristic frequency $\nu_{\rm s}$ and luminosity $L_{\rm s}$, it can be shown
that  $f\simeq {22} \, L_{\rm s, 46} / R_{15} \nu_{\rm s, 16} \delta^3$ 
Substitution of parameter values relevant to the modeling of the 13-day flare, namely $L_{\rm s, 46}=1$, $\nu_{\rm s, 16}=8$, $R_{15}=3.2$ and $\delta=20$, results in $f\approx 10^{-4}$. Interestingly, this is comparable to $f_{\rm EBL}$. Yet, the neutrino luminosity produced via photomeson interactions during the propagation in the ISM,
 $L^{\rm prop}_{\nu}$, is expected to be much lower than that produced internally in the blazar emission region, $L^{\rm int}_{\nu}$. The respective ratio is given  by $L^{\rm prop}_{\nu} / L^{\rm int}_{\nu} \approx f_{\rm EBL} L_{\rm p}^{\rm esc}/L^{\rm int}_{\nu}$, where $L_{\rm p}^{\rm esc}\equiv L_{\rm n}$.
The neutron luminosity is, in turn, given by $L_{\rm n}\simeq (20/6) L_{\nu}^{\rm int}$, where we assumed the production of $\pi^{\pm}$ in a single photopion interaction, leading to  $n : \nu = 1:6$ and that $E_{\rm n} \simeq 20 E_{\nu}$ (see also, \citep{kistler14}).
 Combining all the above we find $L^{\rm prop}_{\nu} / L^{\rm int}_{\nu} \approx (20/6) f_{\rm EBL} \approx 3\times 10^{-4} << 1$.

\section{Summary}
\label{summary}
We presented calculations of the  expected neutrino emission from flaring
periods of the nearby blazar \mrk \ in the context of a one-zone leptohadronic
model for its MW photon emission.  In this scenario, protons are accelerated
and subsequently injected in the emission region of the blazar, where they
pion-produce on the internal synchrotron radiation emitted by a co-accelerated
relativistic electron population.  High-energy neutrinos, which are produced
through the decay of charged mesons, are the final product of photopion
interactions, and may escape the blazar unimpeded, thus providing the smoking
gun for hadron acceleration in blazars.

Using a time-dependent, energy-conserving leptohadronic numerical code
\citep{DMPR2012} we modeled the photon SEDs of the 13-day flare of 2010, which
was the target of an unprecedented MW campaign \citep{aleksic15}; the flaring
episode was simultaneously (within 2 or 3 hours) observed from radio
wavelengths up to the VHE $\gamma$-ray regime.  Based on the model-derived
daily neutrino spectra, we calculated the IceCube muon neutrino event rate of
the 13-day flare, and showed that at energies $>100$~TeV it is comparable to the one
expected  from a longer but non-flaring period of \mrk, i.e. during
quiescence.  We concluded that an accumulation of similar flares over several years would be necessary to produce
a meaningful signal for IceCube. 

The detailed modeling of the 13-day flare revealed a strong correlation of the
expected high-energy neutrino flux with the photon flux in various energy
bands, ranging from soft X-rays ($\sim 2-10$~keV) up to VHE ($>200$~GeV)
$\gamma$-rays. In particular, the relation between the high-energy neutrino and
0.1-300~GeV photon fluxes was found to be $F_\nu \propto F_\gamma^A$ where
$A\sim 1.6$.  We applied the relation, assuming it is valid for longer time
periods as well, to the long-term \fermi \ $\gamma$-ray light curve of \mrk \
that spans $\sim 6.9$ years and coincides with the operation period of the full
IceCube detector. We then estimated the cumulative number of
$\nu_\mu+\bar{\nu}_\mu$ events above 1 PeV for the full IceCube detector
livetime, and found $3.59\pm0.60$ ($2.73\pm0.38$) events with
(without) major flares included in our analysis. This estimate exceeds,
within the uncertainties, the $95\%$ ($90\%$) threshold value for the detection
of at least one neutrino event.  Meanwhile, the most conservative scenario, where no
correlation of $\gamma$-rays and neutrinos is assumed, predicted $1.60\pm0.16$
$\nu_\mu+\bar{\nu_\mu}$ events, still below the 90$\%$ IceCube sensitivity.

In conclusion, by utilizing  model predictions about 
the correlation of PeV-neutrinos and $\gamma$-rays, experiments like IceCube can 
focus their neutrino searches on well-monitored sources, such as \mrk, \ 
{and stack periods where high neutrino activity is expected, such as in  major $\gamma$-ray flares}.

\section*{Acknowledgments}
We thank Prof. E.~Resconi and Prof. A.~Mastichiadis for useful comments on the manuscript. We thank
Dr.~T.~Hovatta for providing us with the \fermi \ light curve.  
{We also thank the anonymous referees for their comments and suggestions.}
M.P. acknowledges support for
this work by NASA through Einstein Postdoctoral Fellowship grant number
PF3~140113 awarded by the Chandra X-ray Center, which is operated by the
Smithsonian Astrophysical Observatory for NASA under contract NAS8-03060.
S.C. acknowledges support by the cluster of excellence "Origin and Structure of the Universe".

\appendix

%

\section[]{Daily SEDs for the period MJD~55266-55276 and model parameter values}
\label{appenA}
In the figures that follow we present the model fits to the daily SEDs for the period  MJD~55266-55276. For clarity reasons, the various emission components have been omitted.
\begin{figure*}
\centering
\subfloat{
\includegraphics[width=0.48\textwidth]{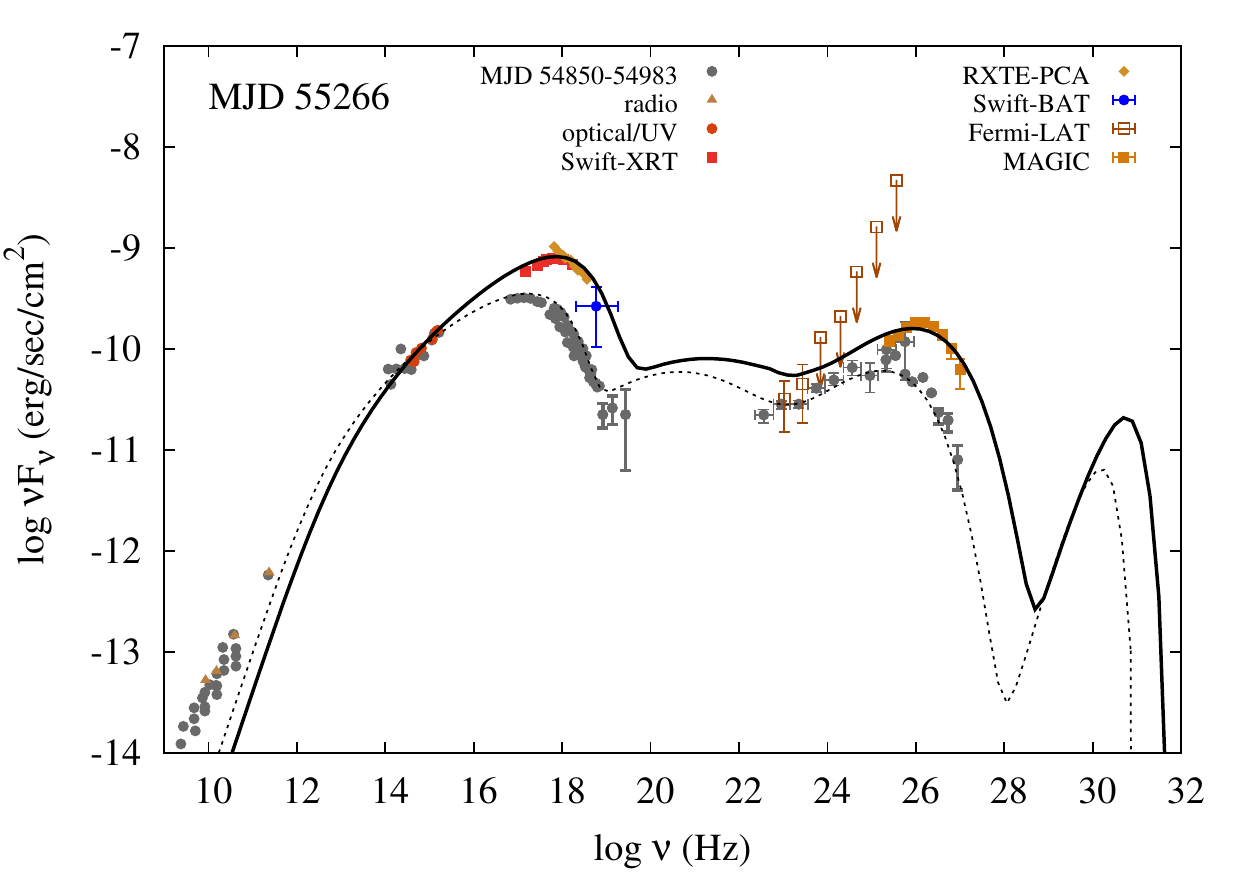}
\includegraphics[width=0.48\textwidth]{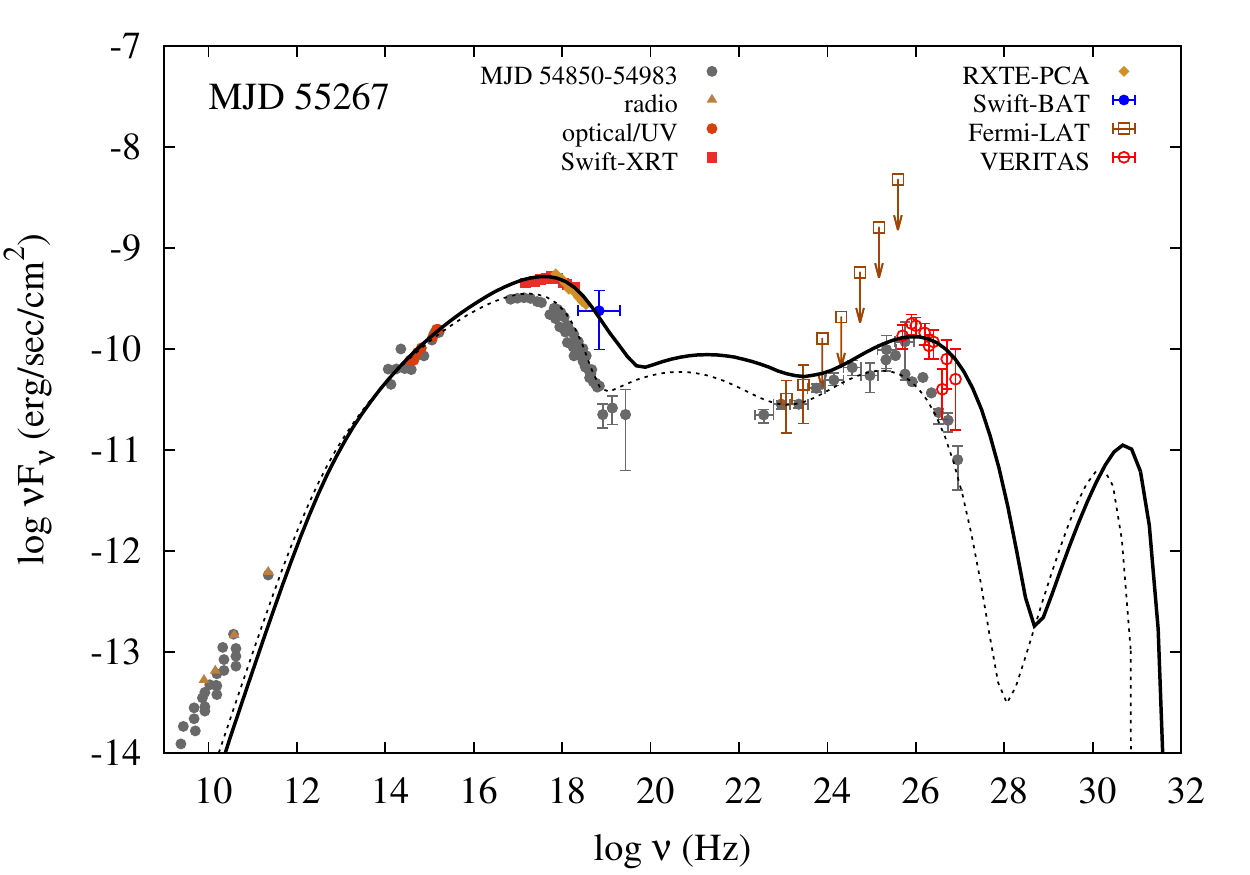}
}

 \subfloat{
 \includegraphics[width=0.48\textwidth]{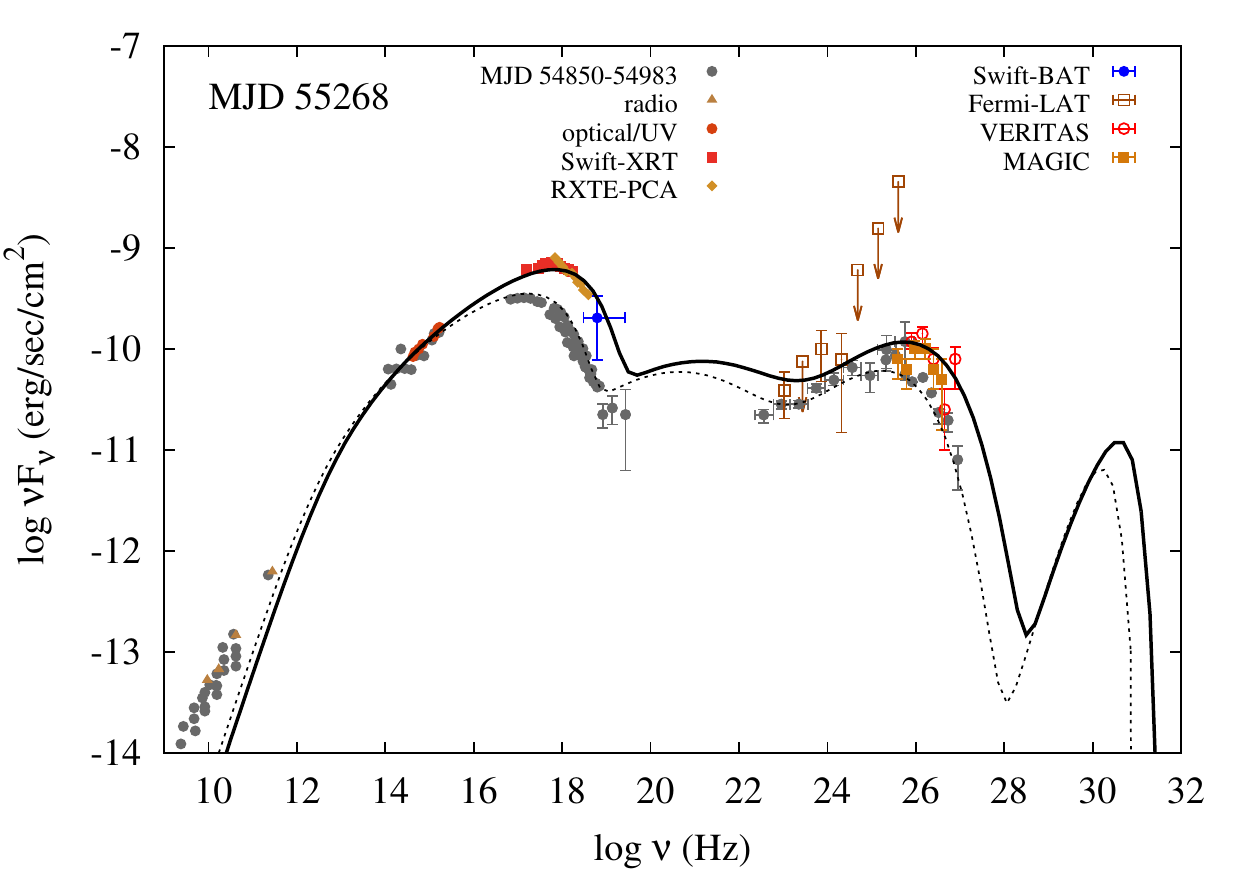}
 \includegraphics[width=0.48\textwidth]{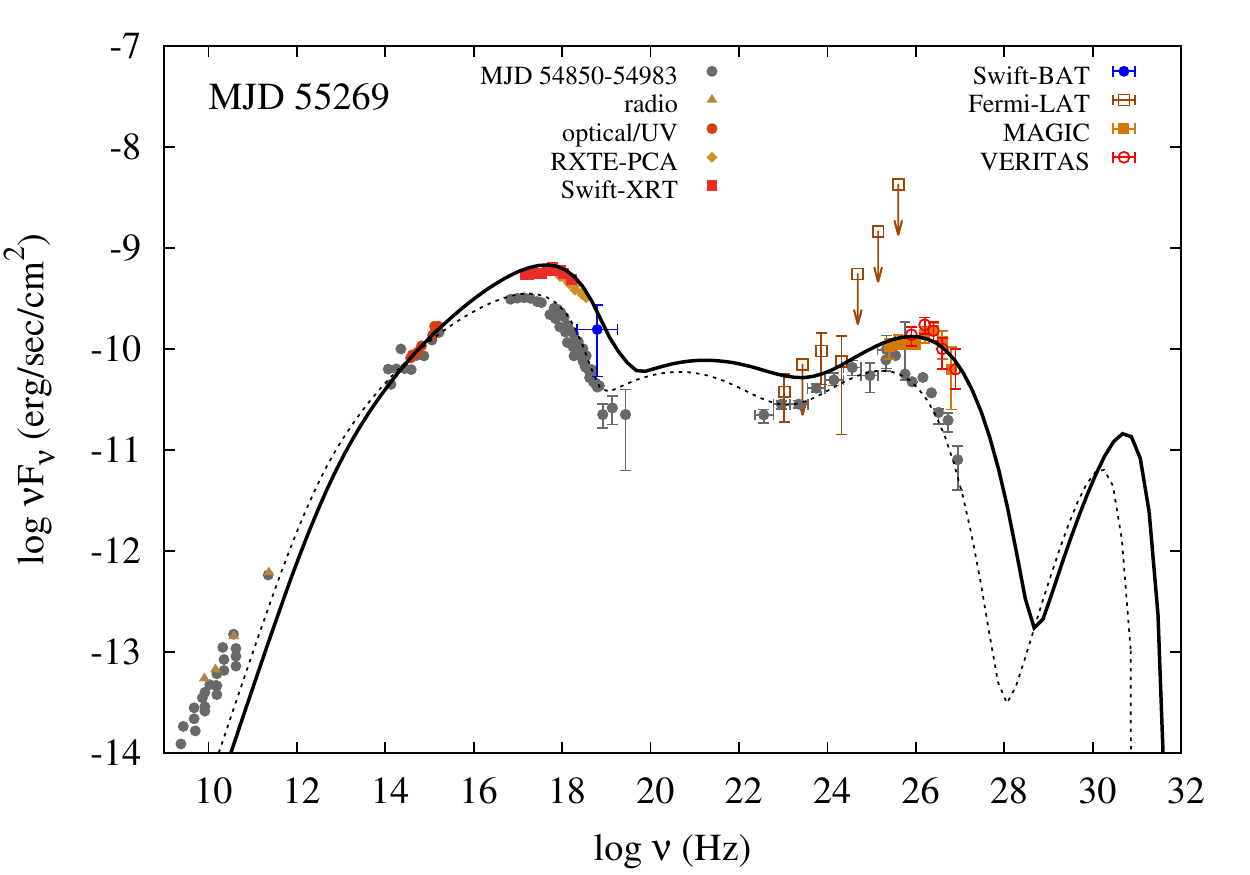}
 }

\subfloat{ 
 \includegraphics[width=0.48\textwidth]{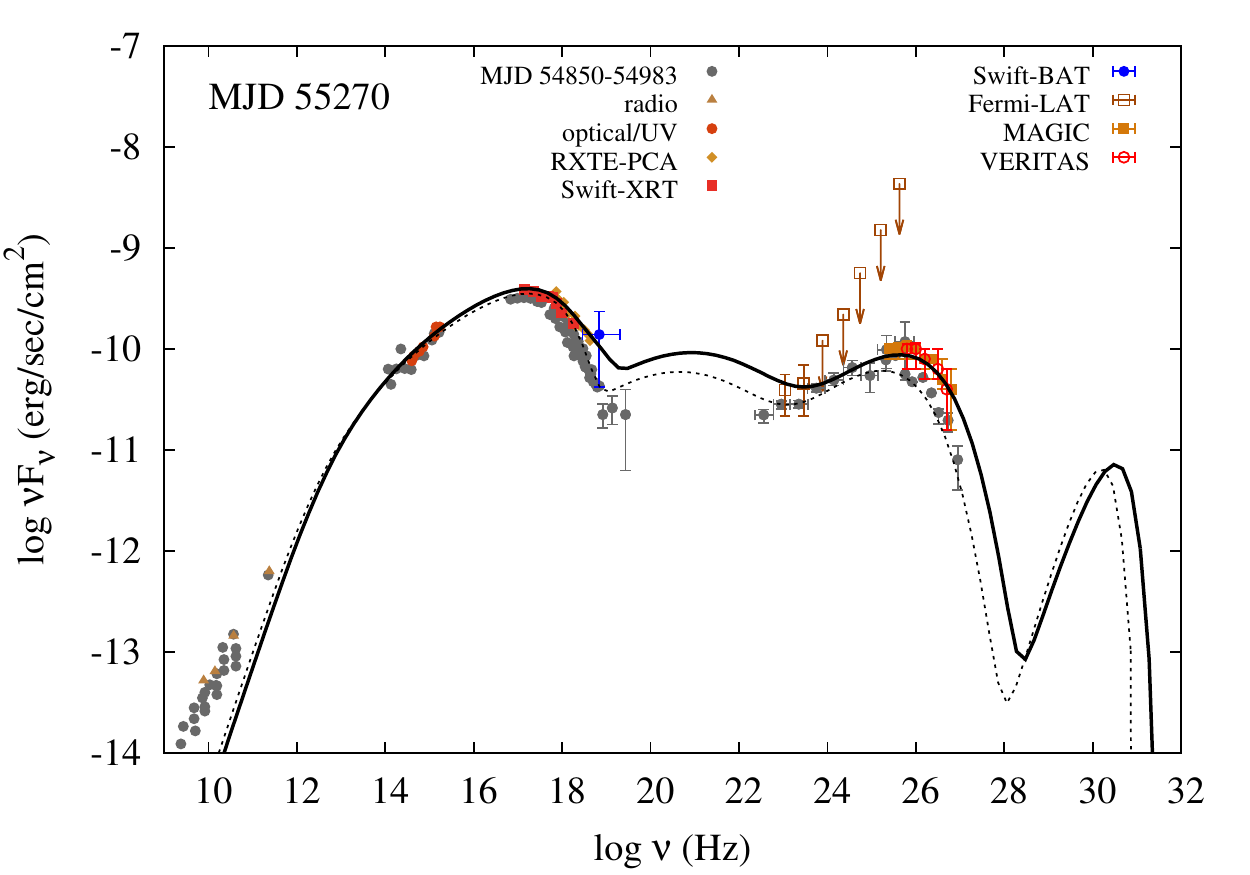}
 \includegraphics[width=0.48\textwidth]{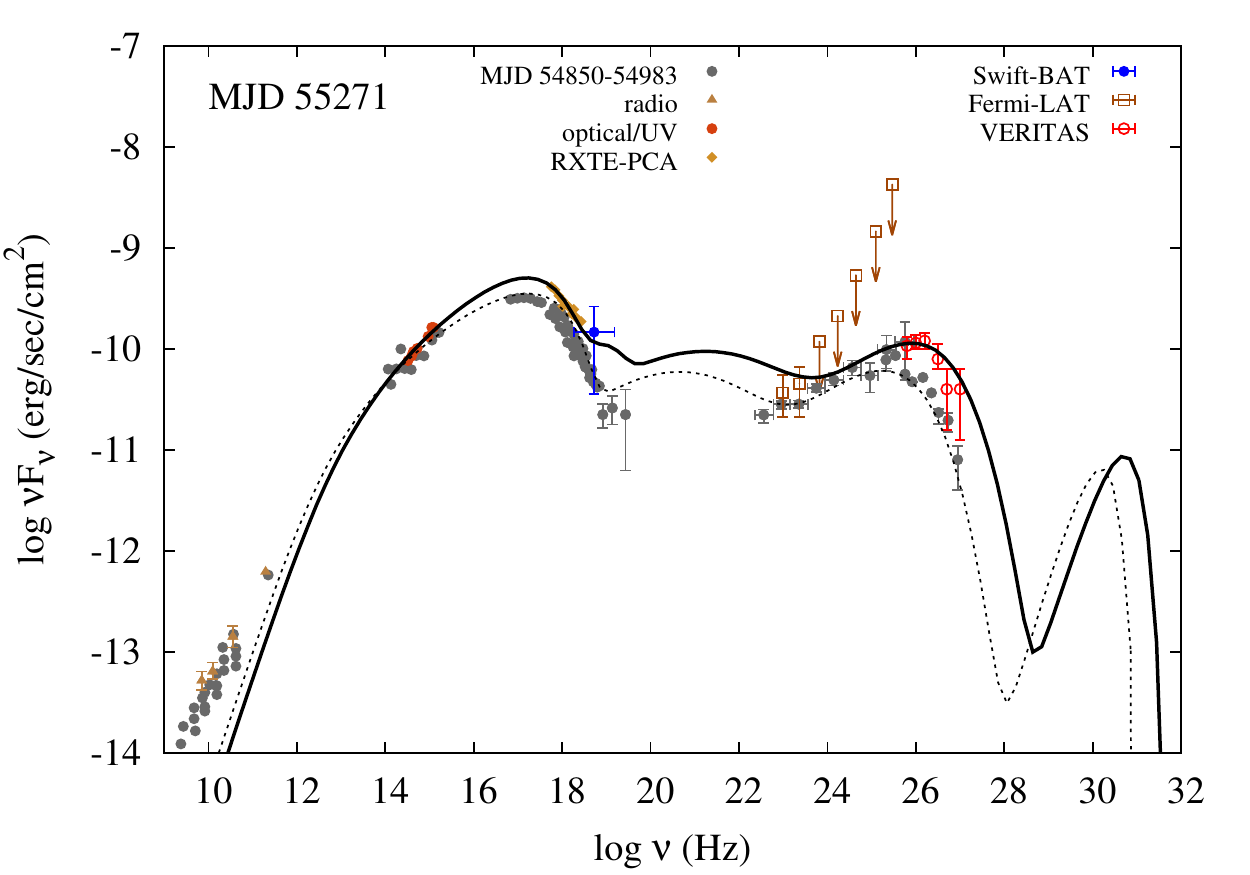}
}
\caption{Model SEDs of \mrk \ for  MJD~55266-55276. All (colored) data-points are from
          \citep{aleksic15}. The grey circles depict the  time-averaged SED of
          \mrk \ over the period MJD~54850-54983 \citep{Abdo2011}. This  is a
          good representation of the blazar  non-flaring (quiescent)
          emission. The model-derived spectra that fit the daily SEDs are
          plotted with  black thick lines. The black dotted line is an indicative fit to
          the quiescent emission.}
 \label{fig2}
\end{figure*}

\begin{figure*}
\ContinuedFloat
\centering
\subfloat{
  \includegraphics[width=0.48\textwidth]{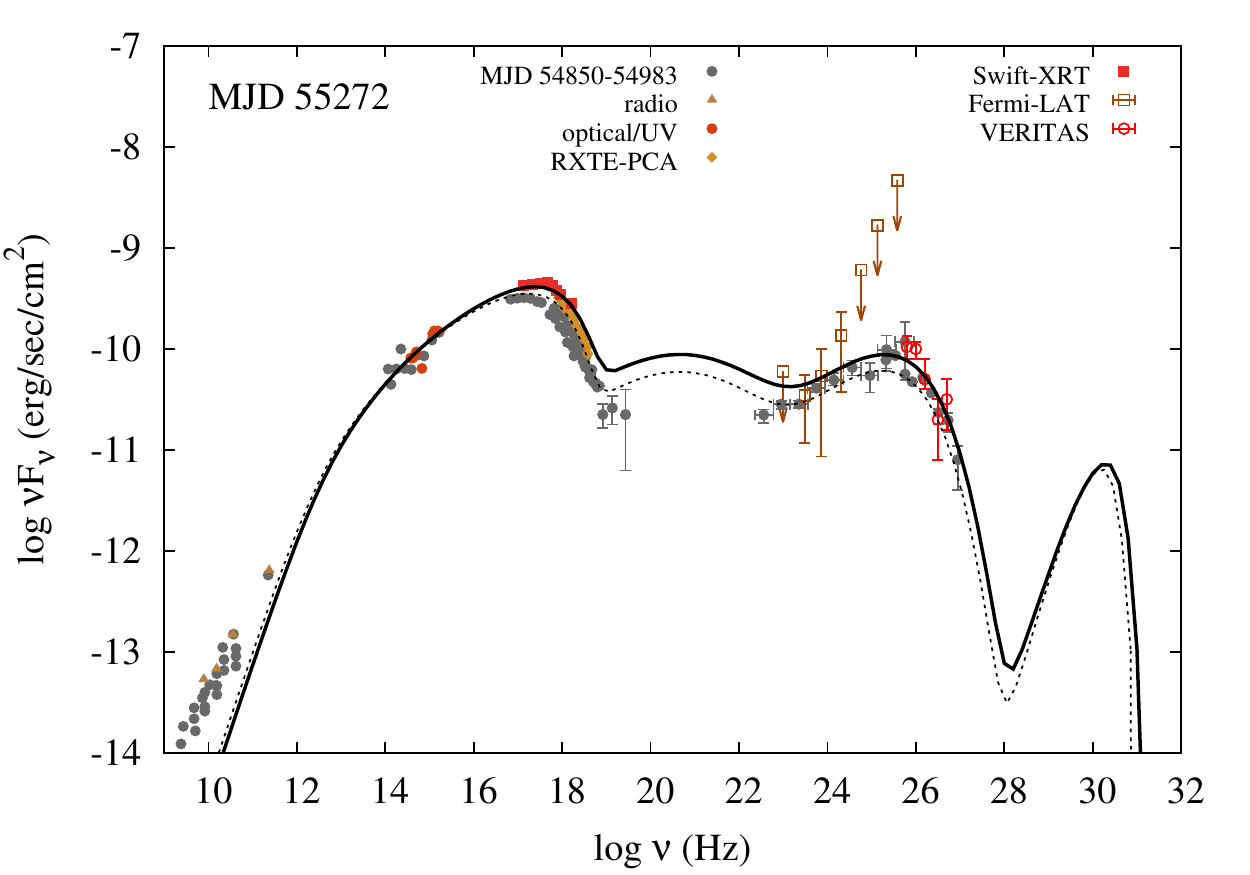}
   \includegraphics[width=0.48\textwidth]{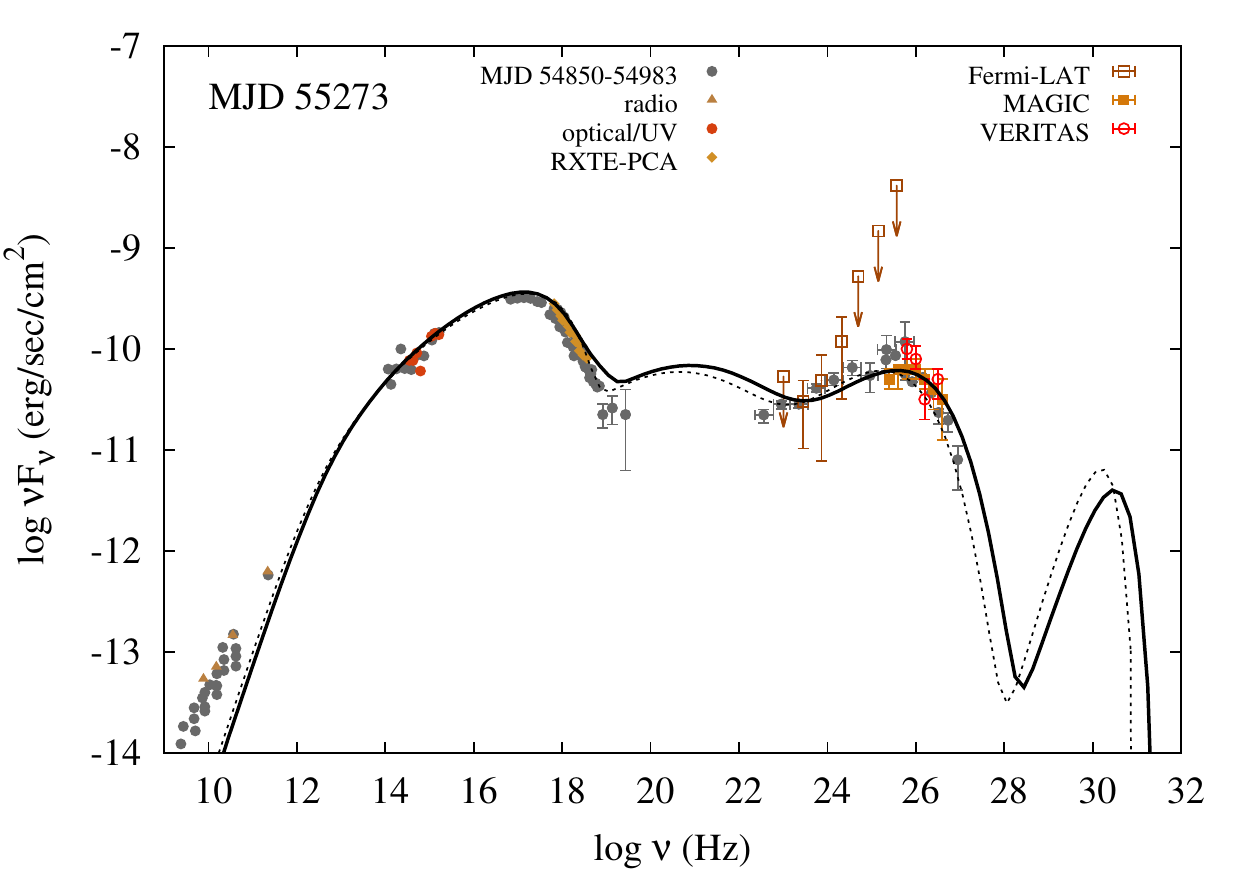}
   }
   
  \subfloat{
  \includegraphics[width=0.48\textwidth]{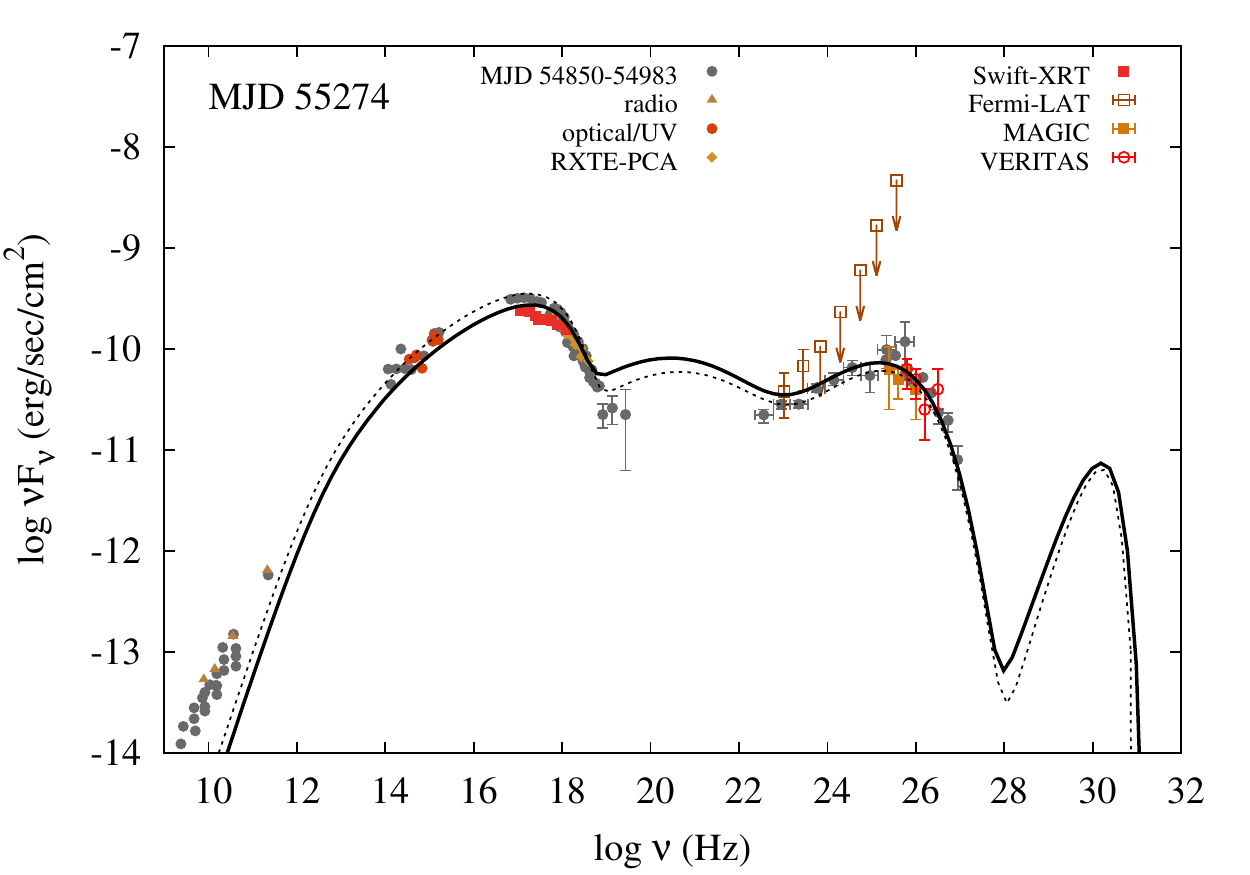} 
  \includegraphics[width=0.48\textwidth]{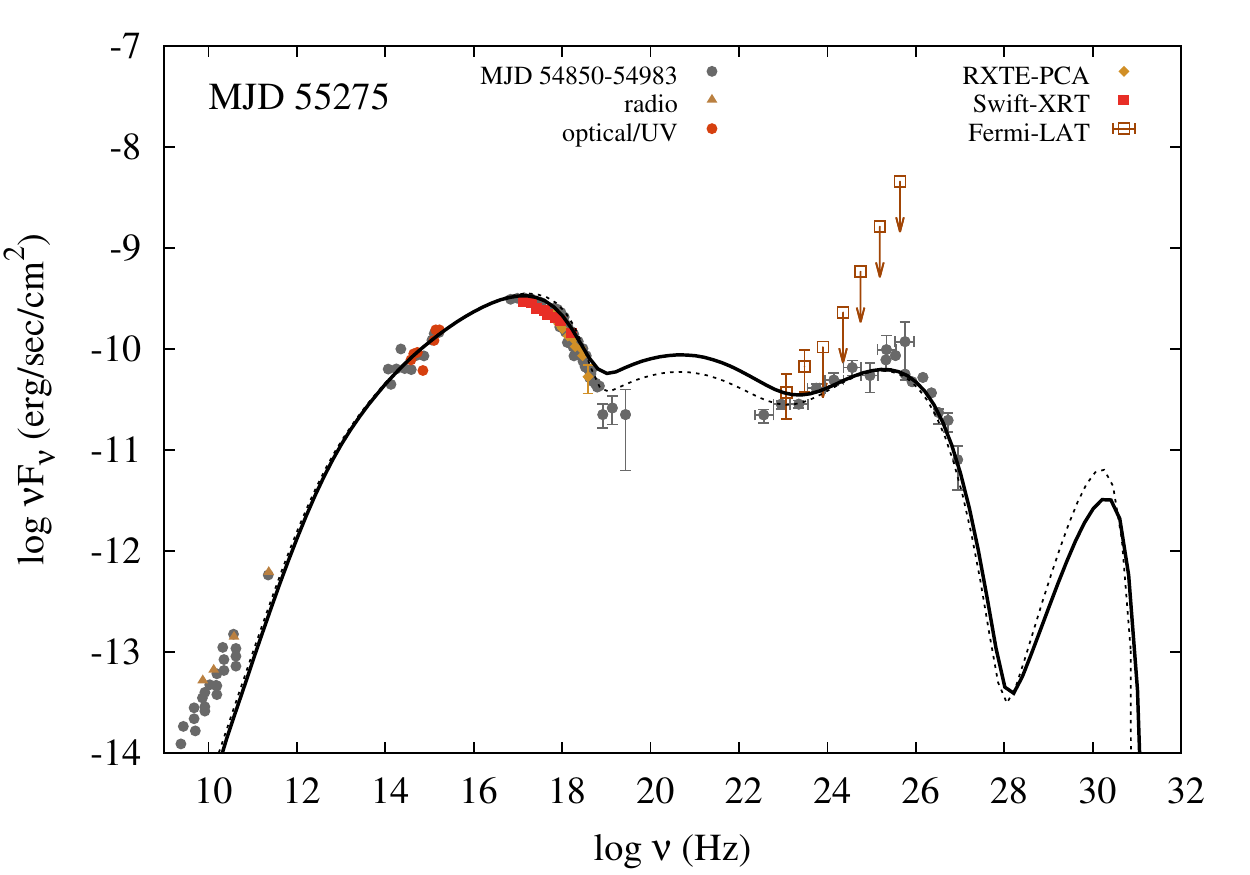} 
  }
  
  \subfloat{
  \includegraphics[width=0.48\textwidth]{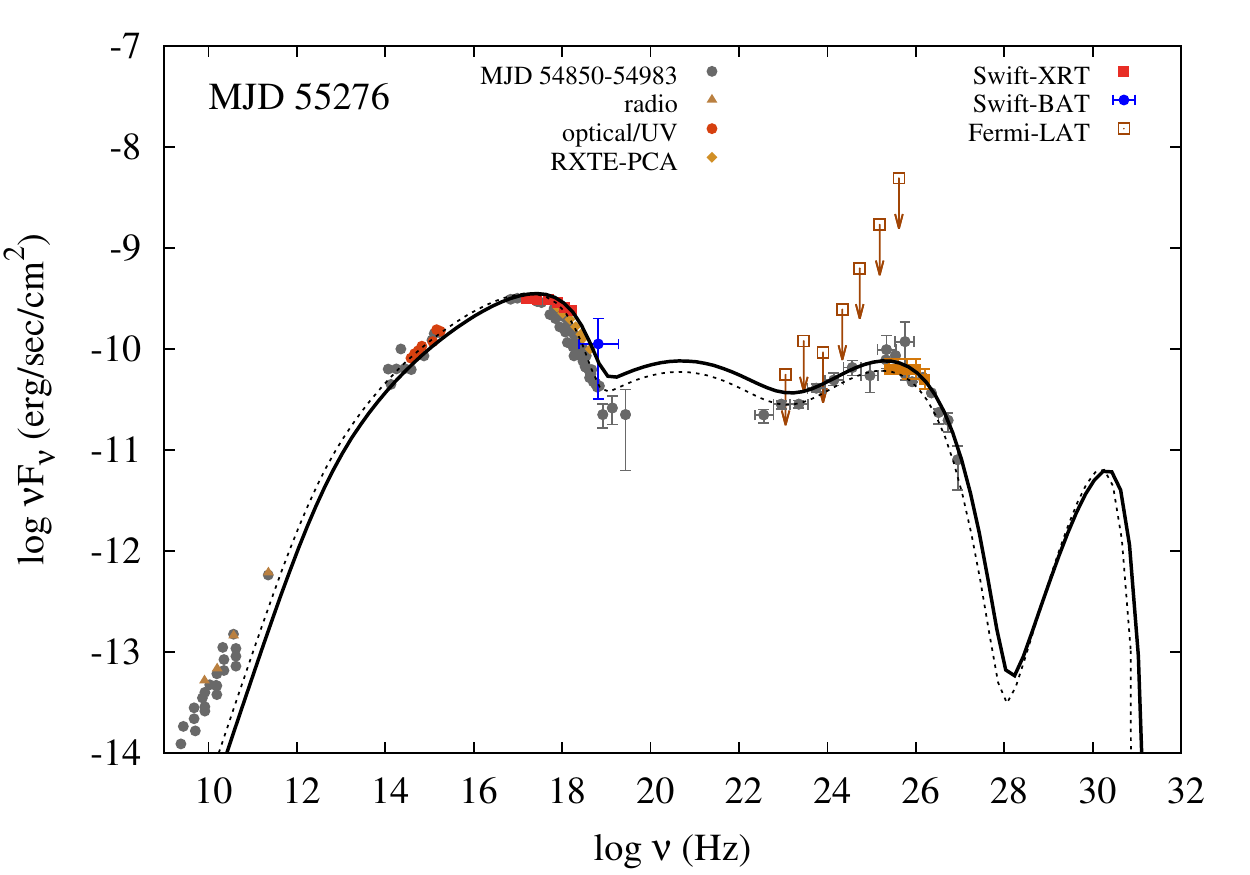}
 } 
 
\caption{Model SEDs continued.}
 \end{figure*}
Table~\ref{table0} summarizes the parameter values of the six model parameters
that were assumed to vary. For completeness, the parameters
used in modeling the SED during the non-flaring period MJD~54850-54983 are also listed.
\begin{table*}
\centering
\caption{Values of the parameters that were allowed to vary while fitting the SEDs from MJD 55265 to MJD 55277. For reference,
the parameter values used for the time-averaged 2009 data \citep{Abdo2011} are also listed.}
\begin{threeparttable}
  \begin{tabular}{c ccc ccc }  
  \hline
Date & $\gemx$ & $\leinj$ & $\gpmx$ & $\lpinj$ & $\doppler$ & $s_{\rm e}$\\
(MJD) & \multicolumn{6}{l}{\phantom{}}\\
\hline
55265& $1.2\times 10^5$ & $2\times 10^{-5}$ &  $8\times 10^6$ & $3.2 \times 10^{-4}$ & 22.3 & 1.2 \\
55266& $10^5$ & $2\times 10^{-5}$ & $6.3 \times 10^6 $ & $4\times 10^{-4}$ & 23 & 1.0 \\
55267& $8 \times 10^4 $& $1.6\times 10^{-5}$ & $6.3 \times 10^6 $ & $5\times 10^{-4}$ & 22.3 & 1.2\\
55268& $10^5$& $2\times 10^{-5}$& $4\times 10^6$ & $5\times 10^{-4}$ & 23.1 & 1.0 \\
55269& $8 \times 10^4 $ & $2\times 10^{-5}$ &	$6.3 \times 10^6 $ & $4\times 10^{-4}$ & 22 & 1.0\\
55270& $5\times 10^4$ & $1.3\times 10^{-5}$& $4\times 10^6$ & $8\times 10^{-4}$ &22 & 1.2 \\
55271& $5\times 10^4$& $2\times 10^{-5}$& $6.3 \times 10^6 $ & $5\times 10^{-4}$ &20.5 & 1.0\\
55272& $6.3 \times 10^4$& $1.6 \times 10^{-5}$& $3.2\times 10^6$ &$10^{-3}$ & 20.8 & 1.2\\
55273& $5\times 10^4$ & $1.6 \times 10^{-5}$ & $4\times 10^6$ & $6.3\times 10^{-4}$ &20.5 & 1.2\\
55274& $5\times 10^4$& $1.1 \times 10^{-5}$& $2.5 \times 10^6$ & $1.6 \times 10^{-3}$ &20 & 1.2\\
55275& $5\times 10^4$&$2\times 10^{-5}$& $3.2 \times 10^6$ &$10^{-3}$ &19 & 1.2\\
55276& $6.3 \times 10^4$ &$1.6 \times 10^{-5}$ & $3.2 \times 10^6$ & $10^{-3}$& 20 & 1.2\\
55277&$5\times 10^4$ &$2\times 10^{-5}$&	$3.2 \times 10^6$ & $10^{-3}$& 19 & 1.2\\
\hline
54850-54983\tnote{a}& $6.3 \times 10^4$ & $1.3\times 10^{-5}$& $3.2 \times 10^6$ &$5.6\times 10^{-4}$ & 21.2 & 1.2\\
 \hline                                                               
  \end{tabular}
  \tnote{a} For the 2009 data, a flatter proton distribution with $s_{\rm p}=0.6$ was adopted.
 \end{threeparttable}
\label{table0}
\end{table*}  
\section[]{Average $\gamma$-ray and neutrino fluxes in quiescence}
\label{appenC}
The average $\gamma$-ray flux during the quiescent period is calculated using eq.~(\ref{Fgamma_q}).
The reported errors were calculated as follows. For every point of the light curve in the time window MJD 54850-54983,
we created a normal distribution of $N=10^5$ random numbers, with mean $\mu=F_{\gamma,i}$ and standard deviation $\sigma=\sigma_{F_{\gamma,i}}$,
where the latter is the statistical error of the measurement.
We then performed  $N$ times the integral in eq.~(\ref{Fgamma_q}) using a five-point Newton-Cotes method for values
drawn from the normal distributions described above. 
The distribution of the average flux $F_\gamma^q$ is shown in Fig.~\ref{fig:hist}. The vertical lines mark the interval where 68\% of the values
lies. These are used to derive the reported errors, namely $F^q_\gamma=4.199^{+0.175}_{-0.165}{\times 10^{-10}}$erg cm$^{-2}$ s$^{-1}$.
\begin{figure}
 \centering
 \includegraphics[width=0.5\textwidth]{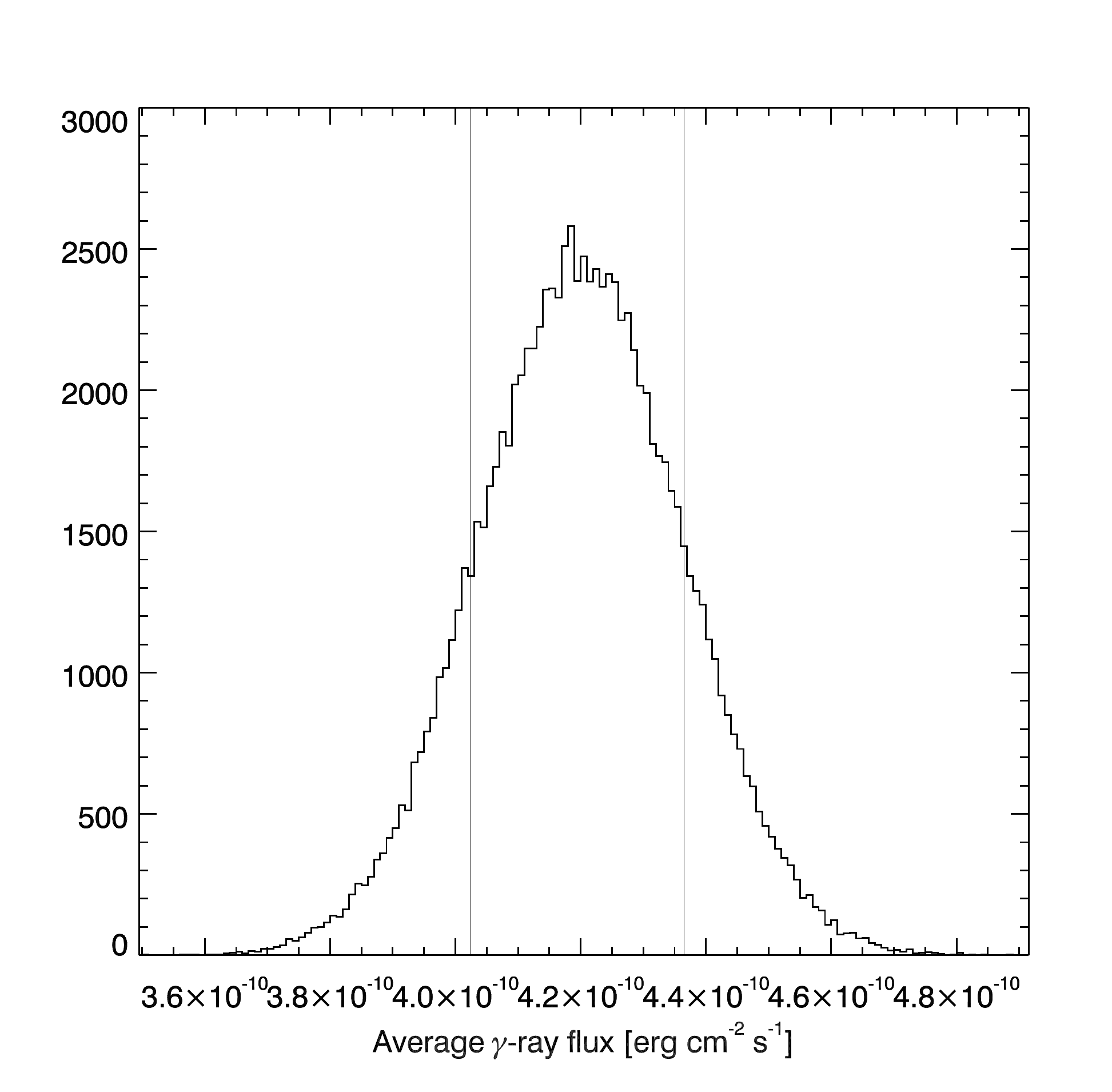}
 \caption{The distribution of the average $\gamma$-ray flux values during quiescence. The vertical lines mark the interval where 68\% of the values
lies.}
 \label{fig:hist}
\end{figure}
A similar procedure is used for the model-predicted quiescent neutrino flux $F_{\nu}^q$, where the values $F_{\nu,i}$
and their respective errors are calculated using eq.~(\ref{linear}).


\bibliographystyle{model1-num-names}
\bibliography{flare}

\end{document}